\documentclass[aps, prl, twocolumn, superscriptaddress, longbibliography]{revtex4-2}
\usepackage[colorlinks, linkcolor=blue, anchorcolor=blue, citecolor=blue]{hyperref}
\usepackage{amsmath}
\usepackage{graphicx}
\usepackage{braket}
\usepackage{multirow}
\usepackage{color}
\usepackage{soul}

\def\cred{\color{red}}
\def\cblue{\color{blue}}

\begin{document}
\title{Optical properties of the infinite-layer La$_{1-x}$Sr$_{x}$NiO$_{2}$
and hidden Hund's physics}
\author{Chang-Jong Kang}
\affiliation{Department of Physics and Astronomy, Rutgers University, Piscataway, New Jersey 08856, USA}

\author{Gabriel Kotliar}
\affiliation{Department of Physics and Astronomy, Rutgers University, Piscataway, New Jersey 08856, USA}
\affiliation{Department of Condensed Matter Physics and Materials Science, Brookhaven National Laboratory, Upton, New York 11973, USA}

\date{\today}

\begin{abstract}
We investigate the optical properties of the normal state of the infinite-layer La$_{1-x}$Sr$_x$NiO$_2$ using DFT+DMFT.
We find a correlated metal which exhibits substantial transfer of spectral weight to high energies relative to the density functional theory.
The correlations are not due to Mott physics,
which would suppress the charge fluctuations and the integrated optical spectral weight as we approach a putative insulating state.
Instead we find the unusual situation, that the integrated optical spectral weight  {\it decreases} with doping and {\it increases } with increasing temperature.
We contrast this  with the coherent component of the optical conductivity,  which
{\it decreases} with increasing temperature as a result of a coherence$-$incoherence crossover.
Our studies reveal that the effective crystal field splitting is dynamical and   increases strongly at low frequency.
This leads to a picture of a Hund's metallic state, where dynamical orbital fluctuations are visible at intermediate
energies, while at low energies a Fermi surface with primarily $d_{x^2 - y^2}$ character emerges.
The infinite-layer nickelates are thus in an intermediate position between the iron based high temperature superconductors where multiorbital Hund's physics dominates,
and a one-band system such as the cuprates.
To capture this physics we propose a low-energy two-band model with atom centered $e_g$ states.
\end{abstract}

\maketitle
\textit{Introduction}---
The recent discovery of superconductivity in the infinite-layer nickelates, Nd$_{1-x}$Sr$_x$NiO$_2$~\cite{Hwang2019}, has attracted intensive interests
due to material similarities with high-$T_{c}$ cuprate superconductors.
Several follow-up experiments confirmed  the superconductivity
\cite{Lee2020-apl,Osada2020-nano,Zeng2020-arXiv,Li2020-arXiv,Gu2020-stm},
with some possibly contradictory observations~\cite{Li2020,Zhou2020}.
Nomura \emph{et al.} estimated the electron-phonon coupling mediated $T_{c}$ to be
$\sim 0.1$ K~\cite{Nomura2019}, much less than the observed $T_{c} \approx 15$ K,
showing the mechanism for superconductivity is unconventional thus
electron correlations play an important role.

There are many experimental investigations into the infinite-layer nickelates
\cite{Hwang2019,Lee2020-apl,Osada2020-nano,Zeng2020-arXiv,Li2020-arXiv,Gu2020-stm,Li2020,
Zhou2020,Lee2020-nm,Mei2019,Goodge2020-arXiv,Crespin1983,Pierre1983,Hayward1999-jacs,
Ikeda2013,Ikeda2016}
and multiple theoretical techniques have been applied
to study their electronic structure
\cite{Nomura2019,Zhang2020-prr,Jiang2019,Norman2020-prx,Pickett2004,Pickett2020-arXiv,
Cano2020-GW,Been2020-arXiv,Hirayama2020,Geisler2020-arXiv,Bernardini2020,
Adhikary2020-arXiv,Sakakibara2019-arXiv,Gao2020-arXiv,Thomale2020,Lechermann2020,
Karp2020-prx,Held2020,Ryee2020,Savrasov2020-prb,Kitatani2020-arXiv,Vishwanath2020,
Chang2019-arXiv,Sawatzky2020-prl,Hu2019-prr,Thomale2020,Zhang2020-Kondo,Gu2020,
Wang2020-arXiv,Lechermann2020-arXiv,YilinWang2020-arXiv,Werner2020-prb,Werner2020-arXiv}.
On the theory side, three different views of these materials are emerging.
In the first one, the infinite-layer nickelate has a cuprate-like correlated $d_{x^2-y^2}$ band near a Mott transition and
an additional uncorrelated ``spectator" band near the Fermi level
which provides self-doping and is
supported by density functional theory (DFT)
\cite{Pickett2004,Nomura2019,Norman2020-prx,Hirayama2020,Zhang2020-prr},
DFT plus dynamical mean-field theory (DFT+DMFT)
\cite{Karp2020-prx,Lechermann2020,Kitatani2020-arXiv},
and model calculations~\cite{Sawatzky2020-prl,Zhang2020-prr,Thomale2020}.
In the second one, it has been suggested that
multiorbital effects are important as, for example, Hund's physics,
using DFT+DMFT~\cite{YilinWang2020-arXiv}, GW+DMFT~\cite{Choi2020-arXiv,Werner2020-arXiv},
and model studies~\cite{Werner2020-prb,Hu2019-prr,Chang2019-arXiv,Vishwanath2020}.
A third approach invokes Kondo physics between correlated and uncorrelated bands.
This is supported by DFT~\cite{Been2020-arXiv}, DFT+Gutzwiller~\cite{Lechermann2020-arXiv},
DFT+DMFT~\cite{Lechermann2020-arXiv,Gu2020},
and model calculations~\cite{Zhang2020-Kondo,Wang2020-arXiv}.
In this paper we present a fourth perspective,
incorporating ideas from the first two viewpoints.
Here frequency renormalization of the crystal fields plays a major role
and results in Hund's multiorbital physics present
at intermediate energies, but hidden at low energies.
This is a new prototype for a strongly correlated metal.

To reach this conclusion we perform fully charge self-consistent DFT+DMFT calculations~\cite{Georges1996,Kotliar2006,Held2007,Haule2010-prb}
implemented in the all-electron full-potential Wien2k package~\cite{Wien2k}
with  the  exact double counting scheme~\cite{Haule2015}
(see computational details in Supplementary Material (SM)~\cite{Suppl}).
This approach was recently shown to give results consistent with
the occupancies measured in high-energy spectroscopies
\cite{Werner2020-arXiv,YilinWang2020-arXiv,Lee2020-nm}.
Here we focus on the basic electronic structure of the infinite-layer nickelates
to extract the basic physics of this class of compounds.
We give special attention to the optical conductivity.
Experiments in this were crucial in identifying
early on the origin and nature of electronic correlations
in different archetypical systems~\cite{Basov2011-rmp}.

\begin{figure}[t]
\centering
\includegraphics[width=0.48\textwidth]{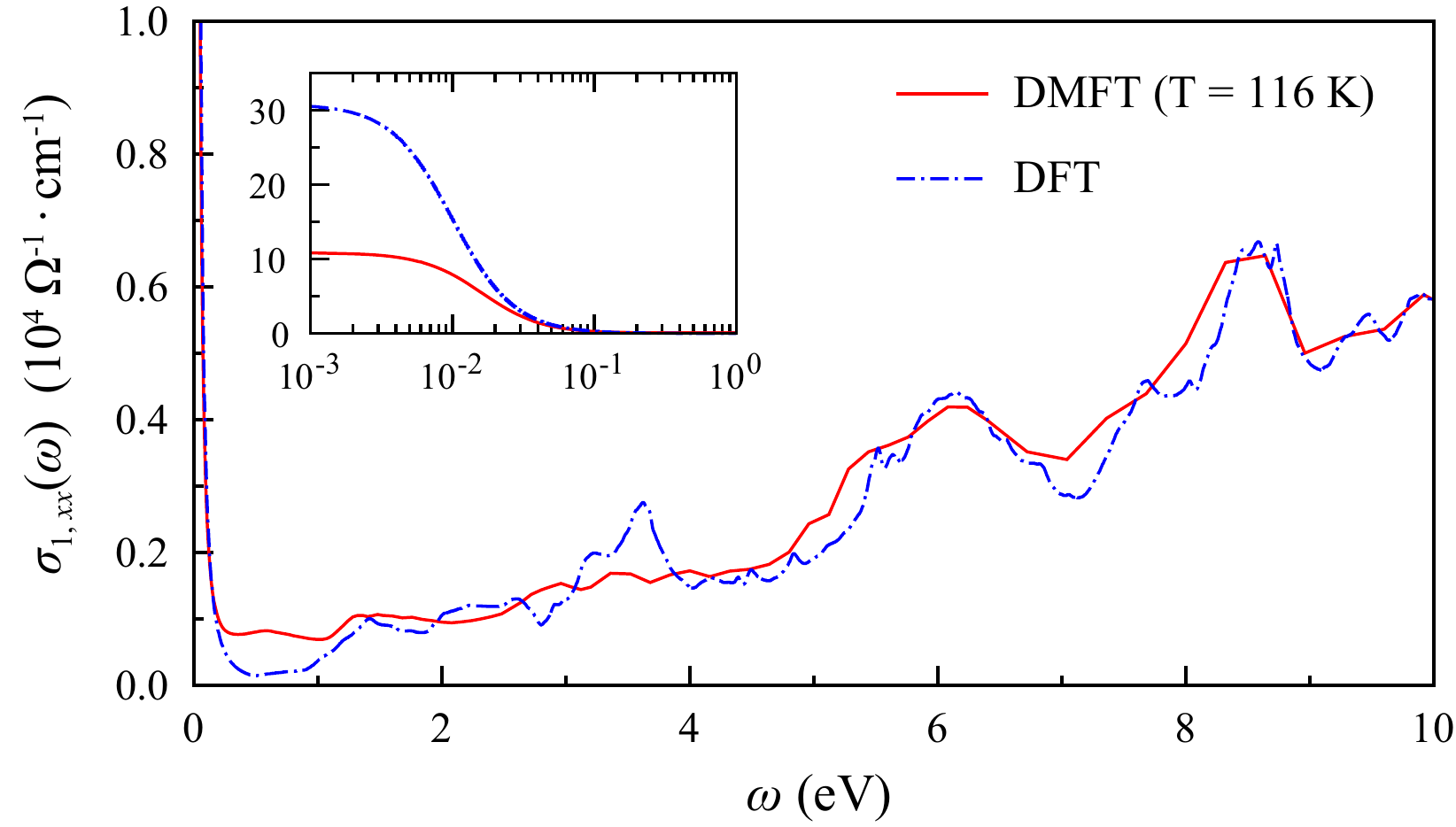}
\caption{
Optical conductivity of LaNiO$_{2}$
in a broad $\omega$ frequency range
computed within both DFT and DFT+DMFT methods.
The inset is provided to magnify the Drude weight.
}
\label{fig:optics}
\end{figure}

\textit{Results: Optical Conductivity}---
The DFT+DMFT optical conductivity is computed with the formalism presented in Refs.~\cite{Haule2010-prb,Suppl} and is shown in Fig.~\ref{fig:optics}
with the DFT reference provided for comparison.
The optical conductivity consists of a Drude weight and interband transitions at $\sim$3.5, $\sim$6, and $\sim$8.5 eV.
The former corresponds to a transition from Ni $3d$ to La $4f$ orbitals and the last two correspond to transitions from O $2p$ to La $4f$ orbitals~\cite{Suppl}.

\begin{figure}[t]
\centering
\includegraphics[width=0.48\textwidth]{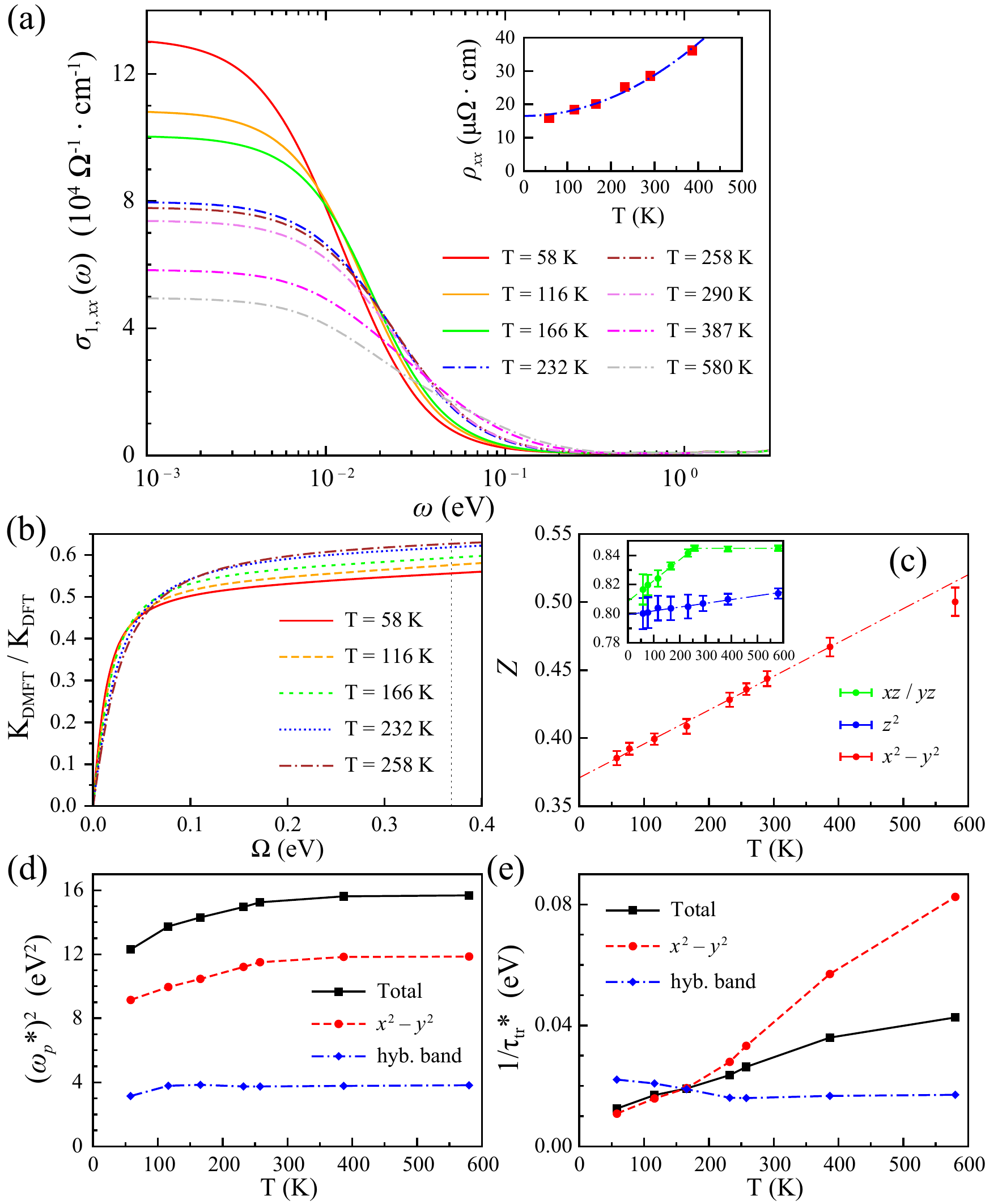}
\caption{
(a) Temperature-dependent optical conductivity of LaNiO$_{2}$ calculated within DFT+DMFT.
The $\omega$-axis is presented in a logarithmic scale.
The calculated resistivity provided in the inset shows the $T^{2}$ behavior.
The blue dash-dotted line in the inset is a guide for the eye by fitting $\rho_{xx}$ to a parabolic function.
(b) The kinetic energy ratio $K_{\text{DMFT}}(\Omega)/K_{\text{DFT}}$
as a function of integration cutoff value $\Omega$ provided for several temperatures.
The vertical dotted line is the kinetic energy integration cutoff
$\Omega_{c}$ = 0.369 eV chosen to exclude a contribution from interband transitions.
(c) Quasiparticle weight $Z$ for Ni $d_{x^2-y^2}$, $d_{z^2}$, and $d_{xz}$/$d_{yz}$ orbitals as a function of temperature.
Error bars originate from the statistical errors in CTQMC simulations.
The dash-dotted lines in (c) are guides for the eye by fitting $Z$ to a linear function.
(d) The effective plasma frequency square $(\omega_{p}^{*})^2$ and
(e) the effective quasiparticle scattering rate $1/{\tau_{\text{tr}}^{*}}$
extracted from the computed optical conductivity within DFT+DMFT
by using the formalism in Ref.~\cite{Deng2014}.
}
\label{fig:optics-T}
\end{figure}

The temperature($T$)-dependent optical conductivity is displayed in Fig.~\ref{fig:optics-T}(a).
The Drude peak develops gradually upon cooling, resulting in a decrease of the resistivity $\rho$ as shown in the inset of Fig.~\ref{fig:optics-T}(a).
The computed $\rho$ follows a $T^{2}$ behavior,
found experimentally at intermediate temperatures~\cite{Ikeda2016}.
At lower temperatures a resistivity upturn below $T \sim 100$ K is observed in experiments~\cite{Ikeda2013,Ikeda2016,Hwang2019}
which we ascribe to disorder effects which are not included in the calculations.

\textit{Results: Integrated Optical Spectral Weight}---
To understand the physics of this material we analyze the integrated spectral weight $K(\Omega) = \int_{0}^{\Omega}\sigma_{1}(\omega)d\omega$ as a function of the cutoff frequency $\Omega$~\cite{Basov2011-rmp}.
Figure~\ref{fig:optics-T}(b) displays $K(\Omega)$ for DMFT
normalized to the DFT Drude weight.
The ratio of $K_{\text{DMFT}}(\Omega)/K_{\text{DFT}}$ decreases upon heating for low cutoff $\Omega$ (less than $\sim$50 meV) as a result of the broadening of the Drude peak.
Above this cutoff, the integrated spectral weight increases with increasing temperature.
The integrated spectral weight up to $\Omega_{c}$ = 0.369 eV
(chosen to exclude a contribution from interband transitions)
in DMFT is about 0.6 of the DFT value at $T$ = 116 K.
Part of the lost weight in the Drude peak is transferred to a low-energy interband transition around $\sim$0.5 eV as shown in Fig.~\ref{fig:optics}.

The reduction of $K_{\text{DMFT}}/K_{\text{DFT}}$
depicted in Fig.~\ref{fig:optics-T}(b)
demonstrates the significance of electronic correlations
which reduces the electronic kinetic energy.
For LaNiO$_{2}$, $K_{\text{DMFT}}/K_{\text{DFT}} = 0.5-0.6$ at $\Omega_{c}$,
thereby suggesting that it is a (moderately) correlated metal.
The kinetic energy ratio is comparable to Hund metal compounds
such as LaFePO and SrRuO$_{3}$~\cite{Basov2011-rmp}.
It is noteworthy that $K_{\text{DMFT}}/K_{\text{DFT}} \approx 0$ for cuprates of La$_{2}$CuO$_{2}$ and Nd$_{2}$CuO$_{4}$, those are charge-transfer insulators, and $\sim$0.2 for La$_{2-x}$Sr$_{x}$CuO$_{2}$ ($x$ = 0.1, 0.15, 0.2)~\cite{Qazilbash2009}.
In addition, in the paramagnetic metallic phase of V$_{2}$O$_{3}$,
which ia a prototypical Mott system,
$K_{\text{DMFT}}/K_{\text{DFT}} \approx 0.2$~\cite{Basov2011-rmp}.
Based on the values of $K_{\text{DMFT}}/K_{\text{DFT}}$,
LaNiO$_{2}$ is far from a Mott system, but close to a Hund's metal.

Notice that the behavior of $K_{\text{DMFT}}/K_{\text{DFT}}$ of LaNiO$_{2}$
as a function of temperature (when $\Omega$ is large)
is the {\it opposite} of what is observed in canonical Mott insulating systems
such as V$_{2}$O$_{3}$
where $K_{\text{DMFT}}(\Omega)$ (or $K_{\text{DMFT}}/K_{\text{DFT}}$) decreases upon heating (within the paramagnetic metallic phase)~\cite{Deng2014}
(see details in SM~\cite{Suppl}).
This reflects the fact that the kinetic energy is reduced
as an insulating state is approached at higher temperatures.
Therefore, LaNiO$_{2}$ is far from a Mott system and closer to a Hund's system
such as BaFe$_{2}$As$_{2}$~\cite{Schafgans2012,U-dependence}.

\textit{Results: Orbital Character}---
We now turn to the orbital character of the different contribution to the optical features.
First we analyze the quasiparticle weight $Z$ as a function of $T$ as depicted in Fig.~\ref{fig:optics-T}(c)~\cite{Z-factor}.
Ni $d_{x^2-y^2}$ has the smallest $Z$
which is drastically smaller than the other $3d$ orbitals having $Z \approx 0.8$.
Notice the strong temperature dependence of $Z$
which increases linearly upon heating,
a clear correlation effect,
which implies the temperature dependence of
the effective mass of the resilient quasiparticles
\cite{Deng2013,Xu2013}.
In contrast, the other orbitals have very weak
or no temperature dependence as shown in the inset of Fig.~\ref{fig:optics-T}(c).
Hence, strong differentiation
between the correlated $d_{x^2-y^2}$ band and
an uncorrelated band with different orbital characters
is present in this system.

The Drude peak could be decomposed into two characters:
the correlated Ni $d_{x^2-y^2}$ and an uncorrelated hybridized band
which includes Ni $d_{z^2}$ and $d_{xz}/d_{yz}$ orbitals~\cite{Suppl}.
It illustrates the multiorbital feature of LaNiO$_{2}$.
The dominant component of the Drude peak is the correlated $d_{x^2-y^2}$
which exhibits strong temperature dependence as shown in $Z$.
The remaining contribution originates from the uncorrelated hybridized band
that are almost temperature independent.
Therefore, $T$-dependent width of the Drude peak is almost solely determined by the electronic correlation exhibited in the $d_{x^2-y^2}$ band.

To gain more insight into the physics of the infinite-layer LaNiO$_{2}$,
we perform a low-energy extended Drude analysis to extract
the effective plasma frequency $\omega_{p}^{*}$
and quasiparticle scattering rate $1/\tau_{\text{tr}}^{*}$
from the computed optical conductivity~\cite{Deng2014}.
In the multiband situation it is useful to
decompose the Drude peak into two contributions,
one coming from the correlated $d_{x^2-y^2}$
and the second from the uncorrelated hybridized band.
The dc conductivity, therefore, can be written as a sum of the two contributions:
$\sigma = \sum_{i} (\omega_{p,i}^{*})^2\tau_{\text{tr},i}^{*}/4\pi$,
where $i$ is a band index.
Figures~\ref{fig:optics-T}(d) and (e) show $(\omega_{p}^{*})^2$ and $1/\tau_{\text{tr}}^{*}$ for each band component as a function of temperature.
The uncorrelated hybridized band shows almost temperature independent
$(\omega_{p}^{*})^2$ and $1/\tau_{\text{tr}}^{*}$.
In contrast, $(\omega_{p}^{*})^2$ for $d_{x^2-y^2}$
shows a linear temperature dependence up to $T \sim 300$ K
where it saturates at the coherence-incoherence crossover,
and was also observed in ruthenates~\cite{Deng2016}
(see Section XII in SM~\cite{Suppl}).
The temperature dependence in $(\omega_{p}^{*})^2$
can be related to the temperature dependence of $Z$ discussed above.
The quasiparticle lifetime $1/\tau_{\text{tr}}^{*}$ for $d_{x^2-y^2}$
is approximately parabolic in temperature below the coherent temperature
and shows a deviation from the quadratic behavior above the coherence-incoherence crossover temperature.

\begin{figure}[t]
\centering
\includegraphics[width=0.48\textwidth]{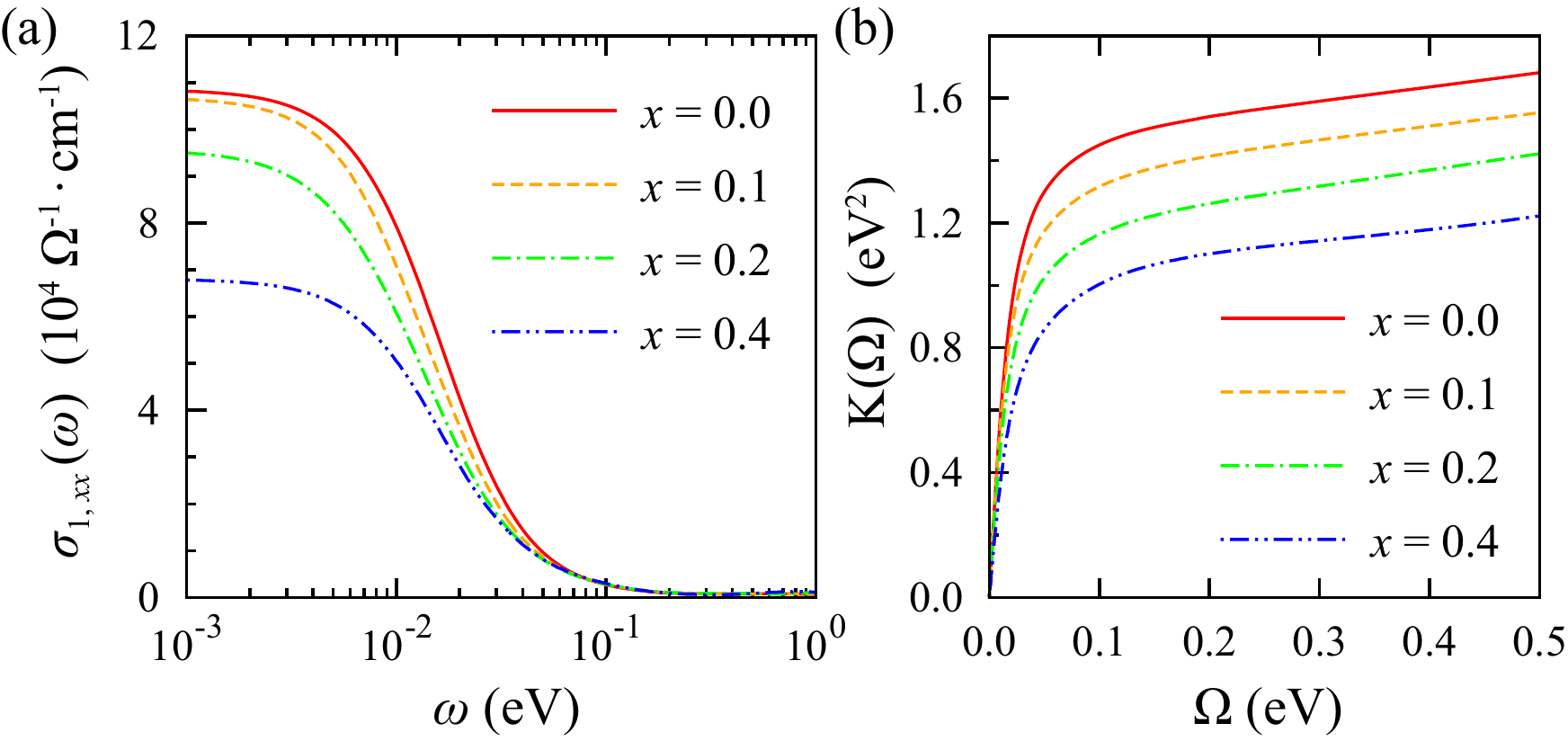}
\caption{
(a) Doping-concentration-dependent optical conductivity of La$_{1-x}$Sr$_{x}$NiO$_{2}$
calculated within DFT+DMFT at $T = 116$ K.
(b) The integrated optical spectral weight $K(\Omega)$ as a function of cutoff frequency $\Omega$ provided for several doping concentration $x$.
}
\label{fig:doping}
\end{figure}

\begin{table}[b]
\centering
\caption{The low-energy carrier number, $n_{e}$, as a function of doping ratio.
$n_{e}$ is defined by the volume of the Fermi surface.
3-dimensional Fermi surface computed with DFT+DMFT is presented in Supplemental Material~\cite{Suppl}.}
\begin{tabular}{p{0.6in}||c|c|c|c|c|c}
\hline
\hline
  & $x=0.0$ & $x=0.1$ & $x=0.2$ & $x=0.3$ & $x=0.4$ & $x=0.5$ \\
\hline
Ni $d_{x^2-y^2}$  & 0.98 & 0.93 & 0.85 & 0.76 & 0.66 & 0.56 \\
\hline
Hybridized band   & 0.07 & 0.03 & 0.02 & 0.01 & $\sim$0.00 & 0.00 \\
\hline
\hline
\end{tabular}
\label{tab:charge_carrier}
\end{table}

\begin{figure*}
\centering
\includegraphics[width=0.95\textwidth]{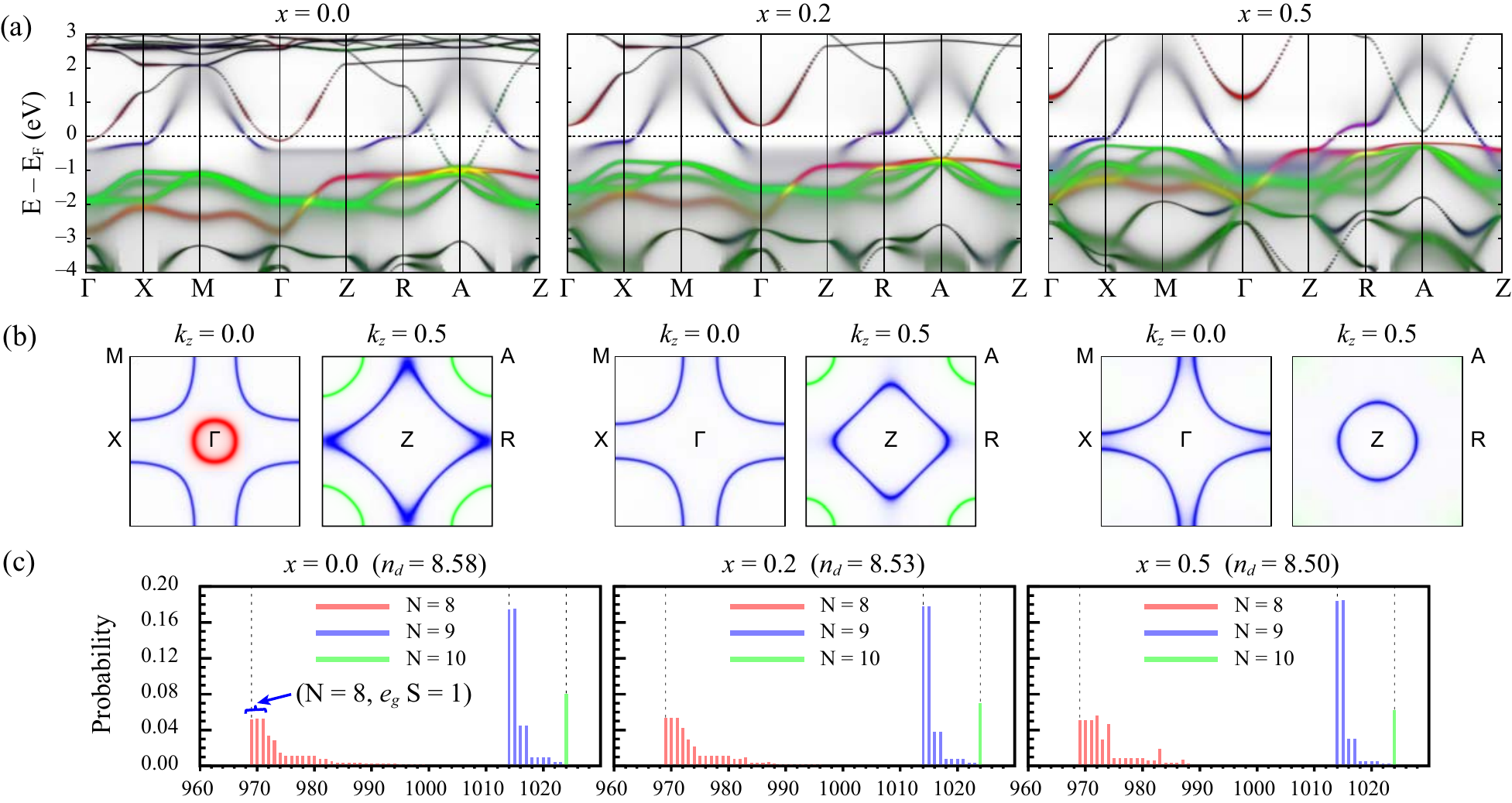}
\caption{
Electronic structures of La$_{1-x}$Sr$_{x}$NiO$_{2}$ computed within DFT+DMFT at $T = 116$ K.
(a) $k$-resolved spectral functions.
(b) Fermi surfaces of La$_{1-x}$Sr$_{x}$NiO$_{2}$ at $k_{z} = 0.0$ and $k_{z} = 0.5$ planes.
Ni $d_{z^2}$, $d_{x^2-y^2}$, and $t_{2g}$ ($d_{xz}/d_{yz}$) orbital characters are presented by red, blue, and green, respectively, in (a) (in (b)).
(c) The DMFT valence histogram of the Ni-$3d$ shell is provided for $x$ = (left) 0.0, (middle) 0.2, and (right) 0.5.
The 1024 possible atomic configurations are sorted by the number of $3d$ electrons of the individual configuration.
}
\label{fig:band}
\end{figure*}

\textit{Results: Doping Dependence}---
Now, we turn our attention to Sr-doped LaNiO$_{2}$,
where Sr doping provides holes to LaNiO$_{2}$.
Figure~\ref{fig:doping}(a) shows the doping-concentration($x$)-dependent optical conductivity of La$_{1-x}$Sr$_{x}$NiO$_{2}$.
The Drude peak and the integrated optical spectral weight $K(\Omega)$
decrease with increasing doping as depicted in Fig.~\ref{fig:doping}.
Since $K(\Omega)$ is proportional to the electronic kinetic energy,
it indicates the {\it decrease} of the kinetic energy upon doping.
This behavior is the opposite of what is observed in a Mott system where
doping increases the electronic kinetic energy~\cite{Imada1998-rmp,Basov2011-rmp}.
This provides further evidence of that
La$_{1-x}$Sr$_{x}$NiO$_{2}$ is far from a Mott transition.
Mott-like behavior was reported in recent GW+DMFT calculations~\cite{Werner2020-arXiv}.

The decrease in the kinetic energy originates from the fact that
the low-energy carrier number, $n_{e}$, defined by the volume of the Fermi surface,
decreases upon doping as presented in Table~\ref{tab:charge_carrier}.
Note that two distinct charge carriers are realized in the Fermi surface,
those are the correlated Ni $d_{x^2-y^2}$ and the uncorrelated hybridized band with Ni $d_{z^2}$
(see Fig.~\ref{fig:band}).
$n_{e}$ for $d_{x^2-y^2}$ decreases significantly upon doping while
$n_{e}$ for the uncorrelated hybridized band is small and varies less.
From the perspective of $n_{e}$, the doped holes mostly
go to the $d_{x^2-y^2}$ band~\cite{occupancy}.

It is important to distinguish the high-energy $d$ occupancy,
which is measured in x-ray spectroscopy~\cite{Lee2020-nm}
and has been shown to be independent of doping~\cite{YilinWang2020-arXiv},
from the low-energy occupancy $n_{e}$
that decreases with increasing hole doping.
This effect, which competes with an increase in $Z$ with increasing hole doping~\cite{YilinWang2020-arXiv},
dominates the behavior of the kinetic energy.

\textit{Results: Electronic Structure}---
The dependence of the $k$-resolved spectral function with doping
is shown in Fig.~\ref{fig:band}(a).
The dominant character at the Fermi level $E_{\text{F}}$
is $d_{x^2-y^2}$ and gives a large Fermi surface (FS)
shown in Fig.~\ref{fig:band}(b).
The uncorrelated hybridized band gives small FSs at $\Gamma$ and $A$.
Hence, the multiorbital character is clearly seen in the calculations
as noticed in earlier works
\cite{Werner2020-prb,Werner2020-arXiv,YilinWang2020-arXiv}.

At $x = 0.2$, the hybridized band with Ni $d_{z^2}$ detaches from $E_{\text{F}}$.
As a result, the FS at $\Gamma$ disappears.
Upon further doping, at $x = 0.5$, another FS from the hybridized band
detaches from $E_{\text{F}}$ as well
and the FS from Ni $d_{x^2-y^2}$ is solely realized.
Therefore, two distinct Lifshitz transitions are realized upon doping~\cite{Savrasov2020-prb,YilinWang2020-arXiv}.

\textit{Results: Two-Band Model}---
The multiorbital character is definitely seen in the DMFT valence histogram
as depicted in Fig.~\ref{fig:band}(c),
where the second largest probability of 0.16 comes from the atomic configuration of ($N=8$, $e_{g}$ $S=1$) with a spin-triplet state within Ni $e_{g}$ states~\cite{CaCuO2}.
Note that ($N=9$, $S=1/2$) with one hole in Ni $d_{x^2-y^2}$ has the largest probability of 0.35.
These two largest probabilities are nearly constant over the doping concentration.
Since the FS has primarily Ni $d_{x^2-y^2}$,
one could interpret it in a one-band scenario as a low-energy model~\cite{Zhang2020-prr,Kitatani2020-arXiv}.
However we find that Hund's coupling $J_{H}$ decreases $Z$
and increases $-\text{Im}\Sigma(i0^+)$ for $d_{x^2-y^2}$
(see Fig.~S14 in SM~\cite{Suppl}),
which is surprising as the atomic ground state configuration has one hole.
This $J_{H}$ dependence of the correlation strength is the hallmark of a Hund's metal
\cite{Haule2009-njp,Medici2011,Georges2013}.
This is because a metallic state requires fluctuations between $d^9$ and $d^8$,
and $J_{H}$ is important in the latter configuration as seen clearly in the valence histogram in Fig.~\ref{fig:band}(c).
We can think of the crystal field as being frequency dependent,
at low energies it leaves $d_{x^2-y^2}$ as the most active orbital,
but at intermediate frequencies both $d_{x^2-y^2}$ and $d_{z^2}$ are important
(see Section XIV and Fig.~S17 in SM~\cite{Suppl}).
Therefore, the infinite-layer nickelate is a Hund's metal
where Hund's correlation is hidden at low energies
but noticeable at intermediate energies.
It is different from the Hund's metal realized in iron pnictides, chalcogenides, and ruthenates,
where a configuration with more than one electron or hole,
makes Hund's correlation prominent in the atomic ground state configuration.
This is made explicit by a two-band Wannier construction
which is atom centered with the symmetry of the two Ni $e_{g}$ orbitals,
but which exhibits the clear difference of $d_{x^2-y^2}$ and $d_{z^2}$
provided in SM
(see Section XIII and Fig.~S16 in SM~\cite{Suppl})
which is different from alternative low-energy Wannier constructions reported before
\cite{Gu2020,Lee2020-nm,Nomura2019,Adhikary2020-arXiv,Been2020-arXiv,Lechermann2020-arXiv,Sakakibara2019-arXiv}.  This perspective provides a qualitative understanding of the variation of
the Hall coefficient with doping and temperature
which is seen in the experiments~\cite{Hwang2019,Li2020-arXiv,Zeng2020-arXiv}
as discussed in SM~\cite{Suppl}.

\textit{Conclusion}---
To summarize, we have computed  the basic electronic structure of
the normal state of La$_{1-x}$Sr$_x$NiO$_2$ within DFT+DMFT with a focus on the temperature and doping dependence of the optical conductivity.
We find signs of strong correlations in the optical response
as the ratio of the optical spectral weight to the DFT band theory
(i.e. $K_{\text{DMFT}}/K_{\text{DFT}}$) is small
(this can be a feature of Mott or Hund's systems).
We find that the evolutions of the optical spectral weight
with (i) temperature and (ii) doping
are qualitatively different from those in a canonical Mott-Hubbard system
and more similar to those of an orbitally differentiated Hund's metal
with a highly correlated $d_{x^2-y^2}$ and a $d_{z^2}$ orbital with weaker correlations.
To reveal the basic physics that govern these materials
we studied the dependence of the physical quantities on the interaction parameters.
We find (1) DMFT valence histogram with enhanced high spin occupation and
(2) strong dependence of the coherence scale with Hund's coupling $J_{H}$,
both are clear signatures of the Hund's metal.
This type of material is quite unique since Hund's physics is hidden at low energies
due to the large crystal field splitting
but noticeable at intermediate energies
where the (dynamical) crystal field splitting
becomes small and Hund's coupling dominates.
It is in an intermediate position between
the multiorbital iron based superconductors
and one-band high-$T_{c}$ cuprate superconductors,
thus opening a new research area in the theory of correlated materials.

\textit{Acknowledgment}---
We are grateful to K. Haule, M. Kim, S. Choi, Y. Wang, H. Miao, and G. L. Pascut for useful discussions.
C.-J. K. and G. K. were supported by the U.S. Department of Energy, Office of Science, Basic Energy Sciences as a part of the Computational Materials Science Program through the Center for Computational Design of Functional Strongly Correlated Materials and Theoretical Spectroscopy.


\renewcommand{\thetable}{S\arabic{table}}
\renewcommand{\thefigure}{S\arabic{figure}}
\renewcommand{\thetable}{S\arabic{table}}
\renewcommand\theequation{S\arabic{equation}}
\setcounter{table}{0}
\setcounter{figure}{0}
\setcounter{equation}{0}
\renewcommand{\bibnumfmt}[1]{[S#1]}
\renewcommand{\citenumfont}[1]{S#1}

\def\cred{\color{red}}
\def\cblue{\color{blue}}

\onecolumngrid

\clearpage

\begin{center}
{\bf \large
{\it Supplemental Material:}\\
Optical properties of the infinite-layer La$_{1-x}$Sr$_{x}$NiO$_{2}$ and hidden Hund's physics
}

\vspace{0.2 cm}

Chang-Jong Kang$^1$ and Gabriel Kotliar$^{1,2}$

\vspace{0.1 cm}

{\small
{\it
$^1$Department of Physics and Astronomy, Rutgers University, Piscataway, New Jersey 08856, USA

$^2$Department of Condensed Matter Physics and Materials Science, Brookhaven National Laboratory, Upton, New York 11973, USA
}
}

\end{center}

\vspace{0.5 cm}

\twocolumngrid

\section{Computational details}

Fully charge self-consistent DFT+DMFT calculations implemented in Wien2k package~\cite{Wien2k}
are performed with formalisms described in Ref.~\cite{Haule2010-prb}
to investigate the infinite-layer LaNiO$_{2}$ system.
The experimental lattice constants of $a = b = 3.871 \text{\AA}$
and $c = 3.375 \text{\AA}$ are adopted for the calculations~\cite{Hayward1999-jacs}.
The muffin-tin radii are 2.50, 1.95, and 1.68 Borh radius for La, Ni, and O, respectively.

We choose a large hybridization energy window from -10 eV to 10 eV with respect to the Fermi level $E_{\text{F}}$
in order to describe both low and high energies physics precisely.
All the five Ni-$3d$ orbitals are considered as correlated ones and
the fully rotational invariant form is applied for a local Coulomb interaction Hamiltonian with on-site Coulomb repulsion $U=5$ eV and Hund's coupling $J_{H}=1$ eV.
The Coulomb parameters $U$ and $J_{H}$ are confirmed by a constraint LDA (cLDA) calculation, where we constructed a 2 $\times$ 2 $\times$ 2 supercell
and put a constrained number of Ni $3d$-electrons, $n_{d} = 9$, in the core state.
It gives the effective Coulomb parameter of $U_{\text{eff}} = F_{\text{eff}}^{0} = U - J = 4.11$ eV, which verifies our $U$ and $J_{H}$ values in the DFT+DMFT calculations.

The continuous time quantum Monte Carlo (CTQMC)~\cite{Haule2007-ctqmc,Werner2006} is adopted for a local impurity solver.
We use generalized gradient approximation (GGA)~\cite{Perdew1996} for the exchange-correlation functional and subtract double counting (DC) term
with the exact DC scheme invented by Haule,
which eliminates the DC issues in correlated materials~\cite{Haule2015}.
The modified Gaussian method~\cite{Haule2010-prb} was used for analytical continuation to obtain the self-energy on the real frequency.
We confirmed that maximum entropy method~\cite{Jarrell1996-maxent} gives the similar self-energy on the real frequency.
The Sr-doping effects are simulated by the virtual crystal approximation.

In the charge self-consistent calculations,
16 $\times$ 16 $\times$ 18 $k$-point mesh is used for the Brillouin zone integration.
In the calculations of the optical conductivity, a more dense $k$-point mesh of 50,000 $k$-points in the full Brillouin zone is adopted.
The broadening factor of 0.01 eV is adopted for DFT+DMFT optical conductivity calculations.
For DFT optical conductivity calculations, the broadening factor of 0.01 eV is used for both intra- and inter-band transitions.
Note that we checked several different broadening factors and found that they do not affect our main conclusions.
The details is provided in Section~\ref{sec:broadening} below.

The electronic structures of both LaNiO$_{2}$ and NdNiO$_{2}$
are nearly identical near $E_{\text{F}}$ when $4f$ electrons in Nd are regarded as core electrons in nonmagnetic fashion (see Section~\ref{sec:NdNiO2}).
Hence, our main conclusions remain unchanged for the low-energy physics
in these rare-earth infinite-layer nickelate materials.

\section{DFT electronic structure}

\begin{figure*}
\centering
\includegraphics[width=0.95\textwidth]{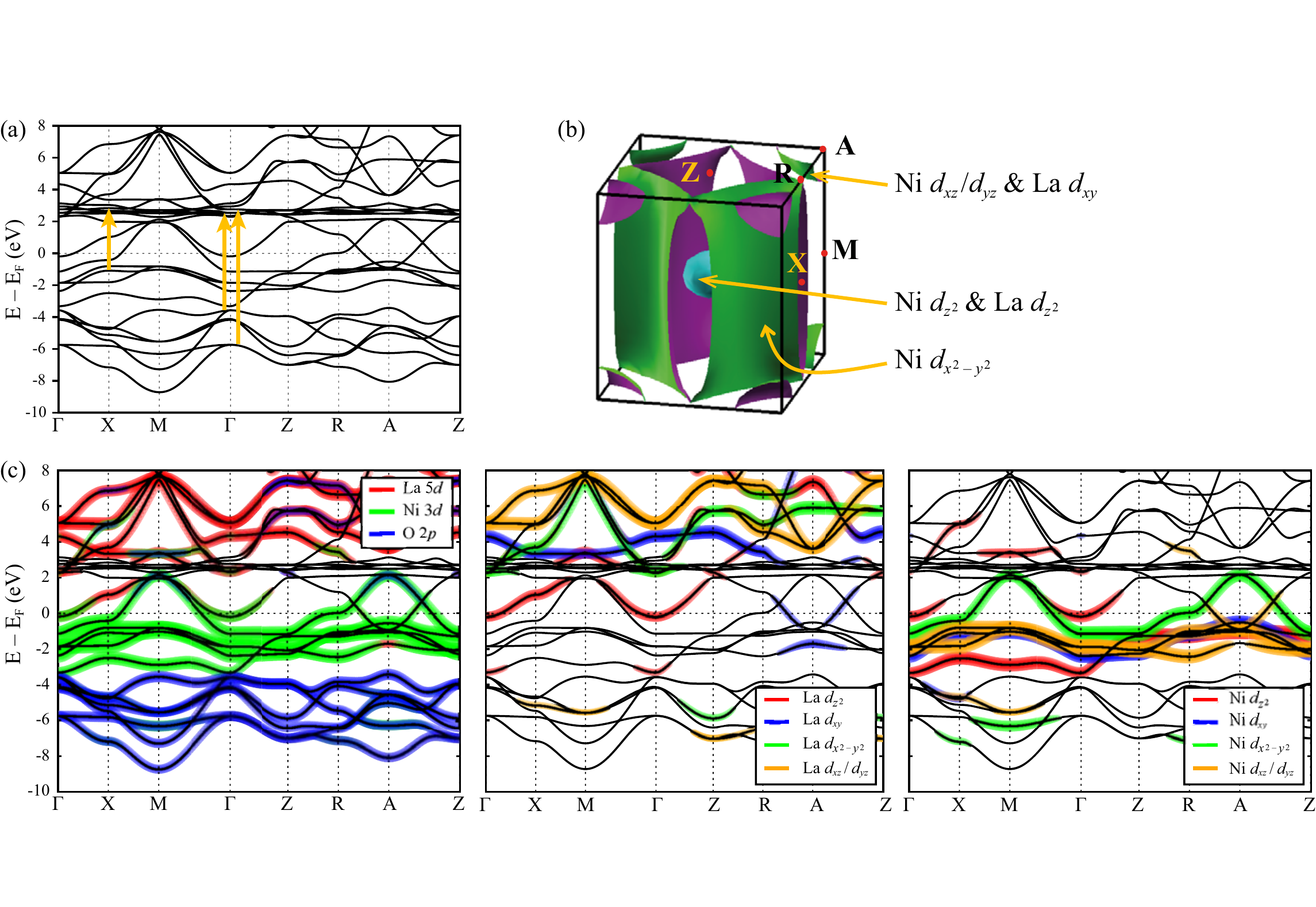}
\caption{DFT electronic structure of the infinite-layer LaNiO$_{2}$.
(a) DFT band dispersion in a broad energy window.
The orange arrows correspond to interband transitions identified in the optical conductivity presented in the main paper.
(b) DFT Fermi surface.
Three Fermi surfaces are realized and their characters are indicated.
(c) DFT band dispersions with orbital characters.
(left) La $5d$, Ni $3d$, and O $2p$
(middle) La $d_{z^2}$, $d_{xy}$, $d_{x^2-y^2}$, and $d_{xz}/d_{yz}$
(right) Ni $d_{z^2}$, $d_{xy}$, $d_{x^2-y^2}$, and $d_{xz}/d_{yz}$ orbital characters are presented.
}
\label{fig:dft-band}
\end{figure*}

Figure~\ref{fig:dft-band} shows the DFT electronic structure of the infinite-layer LaNiO$_{2}$.
Ni $3d$-orbital character is realized in a energy window of (-4 eV, 2 eV), thereby confirming that the large hybridization energy window of (-10 eV, 10 eV) in the DFT+DMFT calculations is sufficiently enough to describe the system.

In Fig.~\ref{fig:dft-band}(a), the orange arrows are provided for the interband transitions observed in the optical conductivity presented in the main text.
Since La $4f$ band locating at $\sim$3 eV has a large density of states,
it is involved in the remarkable interband transitions.

Figure~\ref{fig:dft-band}(c) shows the DFT band dispersion with orbital characters.
Two bands cross $E_{\text{F}}$ in terms of band index,
however more than two orbital characters are realized.
Ni $d_{x^2-y^2}$ crosses $E_{\text{F}}$ and gives a large Fermi surface over the Brillouin zone as presented in Fig.~\ref{fig:dft-band}(b).
Ni $d_{z^2}$ and $d_{xz}/d_{yz}$ cross $E_{\text{F}}$ as well and hybridize significantly with La $d_{z^2}$ and $d_{xy}$, respectively.
They give sphere-like Fermi surfaces at $\Gamma$ and $A$, respectively, as shown in Fig.~\ref{fig:dft-band}(b).

The DFT+DMFT calculations identify the same orbital characters crossing $E_{\text{F}}$.
Hence, the infinite-layer LaNiO$_{2}$ shows the multiorbital character apparently.

\section{Decomposition of the Drude peak}
Since the two bands cross $E_{\text{F}}$,
they contribute the intraband transition in the optical conductivity.
In the DFT+DMFT method, the real part of the optical conductivity
is computed as follows~\cite{Haule2010-prb}:
\begin{eqnarray}
\sigma_{1}^{\mu\nu}&(\omega)&  \nonumber
\\
& = & \frac{\pi e^2}{V_{0}}\sum_{\textbf{k}}
\int d\omega (-\frac{\partial f}{\partial\omega})
\text{Tr}[v_{\mu}(\textbf{k})A(\textbf{k},\omega)v_{\nu}(\textbf{k})A(\textbf{k},\omega)],
\nonumber
\\
\label{eq:optics}
\end{eqnarray}
where $e$, $V_{0}$, $f$, $v_{\mu}(\textbf{k})$, and $A(\textbf{k},\omega)$ are the elementary charge, the volume of the unit cell, the Fermi-Dirac distribution function, the Fermi velocity along $\mu$-direction, and the spectral function, respectively.

\begin{figure*}
\centering
\includegraphics[width=0.6\textwidth]{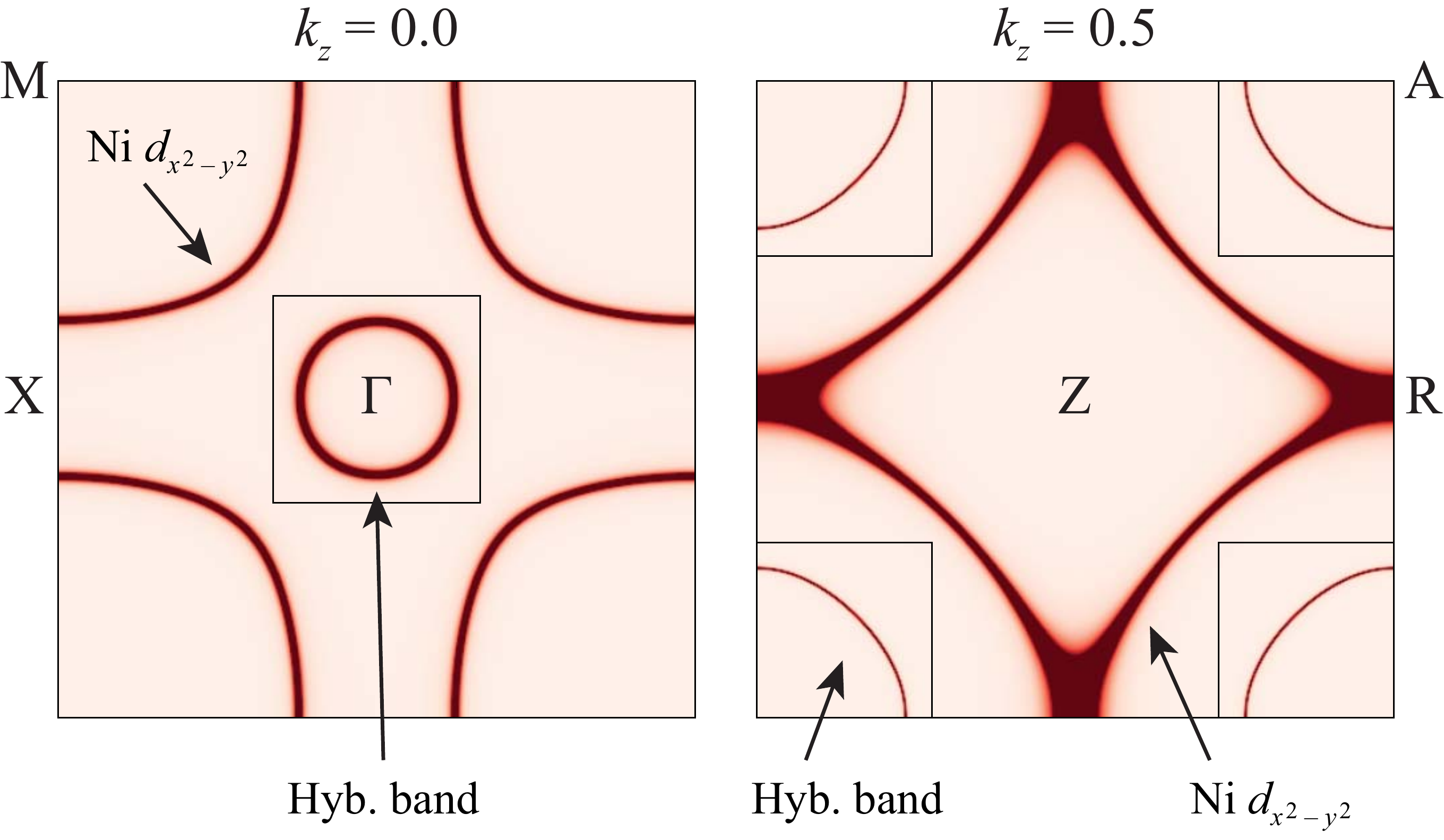}
\caption{
Discretization of the Brillouin zone so as to decompose the Drude peak into two band components (Ni $d_{x^2-y^2}$ and an uncorrelated hybridized band)
that cross the Fermi level.
In the Fermi surface of LaNiO$_{2}$ computed with DFT+DMFT, each band character is well separated spatially in the Brillouin zone indicated by the black squares.
}
\label{fig:BZ}
\end{figure*}

Since Eq.~(\ref{eq:optics}) contains the $k$-sum in the Brillouin zone,
the Drude peak could be decomposed into each band character if
the all band characters are well separated spatially in the Brillouin zone.
Figure~\ref{fig:BZ} demonstrates that all distinct band characters
of the Fermi surface are well separated in the Brillouin zone.
Note that the hybridized band has multiple orbital characters
and gives two Fermi surfaces at $\Gamma$ and $A$.
Therefore, the low-energy Drude peak could be decomposed into two band components:
the correlated Ni $d_{x^2-y^2}$ band and the remaining uncorrelated hybridized band.
Then, decomposition of the Drude peak is accomplished by performing the $k$-sum in each discretized Brillouin zone indicated in Fig.~\ref{fig:BZ}.

\begin{figure*}
\centering
\includegraphics[width=0.98\textwidth]{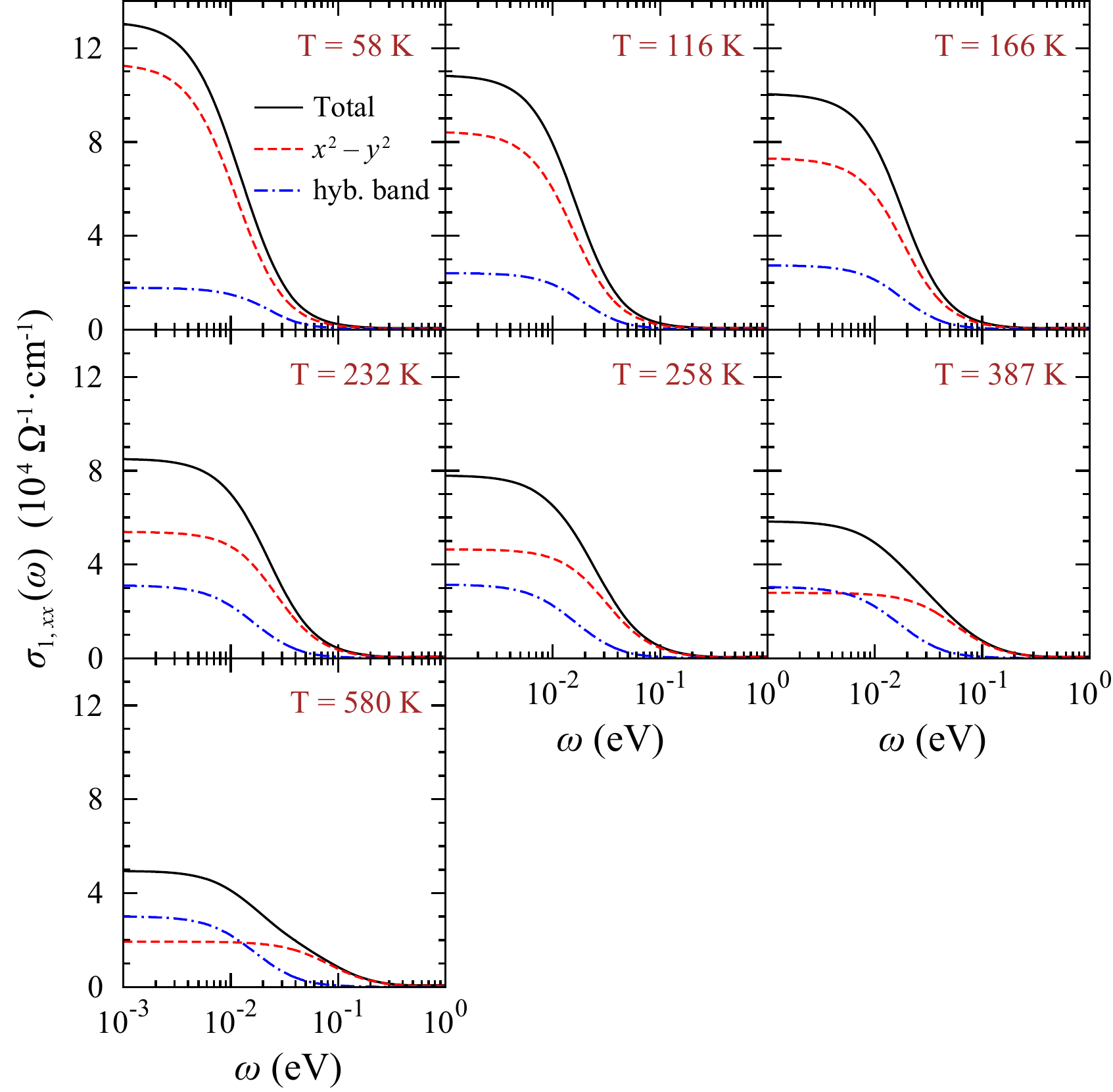}
\caption{Decomposition of the Drude peak in the infinite-layer LaNiO$_{2}$ at a broad temperature range.
The Drude peak computed with DFT+DMFT is decomposed into each band component that crosses the Fermi level,
that are Ni $d_{x^2-y^2}$ and an uncorrelated hybridized band.
}
\label{fig:Drude}
\end{figure*}

Figure~\ref{fig:Drude} shows the decomposition of the Drude peak
computed within DFT+DMFT at a broad temperature range.
The effective plasma frequency $(\omega_{p}^{*})^2$
and the effective scattering rate $1/\tau_{\text{tr}}^{*}$
for each band component are extracted from the data presented in Fig.~\ref{fig:Drude}
and they are depicted as a function of temperature in the main text.

\section{Impact of broadening factor on optic calculations}
\label{sec:broadening}

Equation~(\ref{eq:optics}) includes the spectral function $A(\textbf{k},\omega)$ that is given by
\begin{equation}
A(\textbf{k},\omega) = \frac{1}{2\pi i} \big( G^{\dagger}(\textbf{k},\omega) - G(\textbf{k},\omega) \big).
\end{equation}
The Green's function $G(\textbf{k},\omega)$ can be expressed as
\begin{equation}
G(\textbf{k},\omega) = \frac{1}{\omega + i\delta + \mu - \varepsilon_{\text{DFT}}(\textbf{k})-\Sigma(\omega)},
\end{equation}
where $\delta$, $\mu$, $\varepsilon_{\text{DFT}}(\textbf{k})$, and $\Sigma(\omega)$ are the broadening factor, chemical potential, DFT energy eigenvalue, and self-energy, respectively.
Therefore, the optical conductivity depends on the broadening factor $\delta$.
Note that there is additional broadening factor $\delta_{c}$
for the correlated bands in addition to the self-energy.
In the calculations, we set $\delta_{c} = 0$.

\begin{figure*}
\centering
\includegraphics[width=0.95\textwidth]{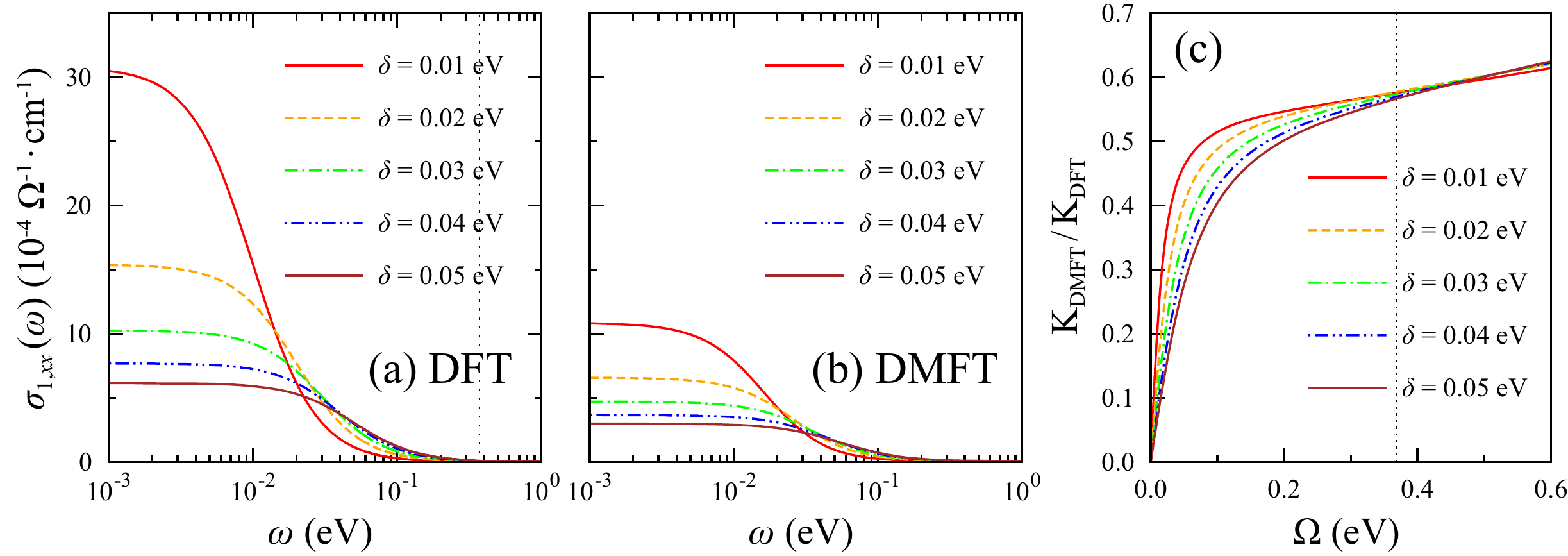}
\caption{Optical conductivity of LaNiO$_{2}$ for several different broadening factors $\delta$.
Optical conductivity computed within (a) DFT and (b) DFT+DMFT methods.
The same broadening factor $\delta$ is used for both intra- and inter-band transitions.
(c) The kinetic energy ratio $K_{\text{DMFT}}(\Omega)/K_{\text{DFT}}$ as a function of integration cutoff value $\Omega$ provided for several different broadening factors.
The vertical dotted line is the kinetic energy integration cutoff
$\Omega_{c}$ = 0.369 eV
chosen to exclude a contribution from interband transitions.
}
\label{fig:broadening}
\end{figure*}

Figures~\ref{fig:broadening}(a) and (b) show the optical conductivity
for several different broadening factors, $\delta$,
computed within DFT and DFT+DMFT methods, respectively.
Even though $\delta$ changes the spectral shape of the Drude component a lot,
it hardly affects the kinetic energy ratio
$K_{\text{DMFT}}(\Omega)/K_{\text{DFT}}$ at the cutoff $\Omega_{c}$ = 0.369 eV,
chosen to exclude a contribution from interband transitions,
as demonstrated in Fig.~\ref{fig:broadening}(c).
An extremely small $\delta$ is not adequate for a numerical reason,
hence $\delta$ = 0.01 eV is the reasonable broadening factor in the calculations.

\section{Comparison with N\lowercase{d}N\lowercase{i}O$_{2}$}
\label{sec:NdNiO2}

There is an issue of the effects of Nd $4f$ electrons in the infinite-layer nickelate NdNiO$_{2}$ \cite{Pickett2020}.
However, the effects only appear at extremely low energies
and are not the subject of this study even though they give rise to interesting physics at very low energies, or at ultralow temperatures.

When the $4f$ electrons in Nd is regarded as core electrons in nonmagnetic fashion, the electronic structures of both LaNiO$_{2}$ and NdNiO$_{2}$ are nearly identical near the Fermi level as demonstrated in Fig.~\ref{fig:ndnio2-dmft}.
Lattice constants for these materials are very similar: $a$ = $b$ = 3.871 ${\AA}$, $c$ = 3.375 ${\AA}$ for LaNiO$_{2}$ and $a$ = $b$ = 3.914 ${\AA}$, $c$ = 3.239 ${\AA}$ for NdNiO$_{2}$. The main effect of the rare-earth Nd element at high temperatures is to have localized $4f$ electrons that could be treated as core electrons.

Figure~\ref{fig:ndnio2-dmft} demonstrates that the different rare-earth elements are not relevant to the main properties of the electronic structure of these infinite-layer nickelate materials, and for these, the $4f$ electrons can be considered core-like, and we show that the core-like electron behaves essentially like La which has no $4f$ electrons, from the perspective of the valence electrons.

Figure~\ref{fig:ndnio2-optics} shows the optical conductivities of LaNiO$_{2}$ and NdNiO$_{2}$ computed based on the electronic structures shown in Fig.~\ref{fig:ndnio2-dmft}.
Note that La $4f$ electrons are treated as valence electrons, however Nd $4f$ as core in the calculations. Therefore, the optical conductivity of NdNiO$_{2}$ does not show any optical transitions related to the Nd $4f$ within the $\omega$ frequency displayed in Fig.~\ref{fig:ndnio2-optics}.
It leads for the two materials to show different optical conductivities in a broad $\omega$ frequency range.
However, as shown in the inset of Fig.~\ref{fig:ndnio2-optics},
both materials have the Drude components that are almost identical.
It results in the similar kinetic energy ratio $K_{\text{DMFT}}/K_{\text{DFT}}$ between the two materials.
Hence, our main conclusions remain unchanged for the low-energy physics in these infinite-layer nickelate materials.

\section{Optical conductivity for different Coulomb U parameters}
We compute the optical conductivity for some different Coulomb repulsion $U$ values.
First, we provide the DMFT valence histogram for some different $U$ values in Fig.~\ref{fig:histogram_U}.
As shown in Fig.~\ref{fig:histogram_U},
an atomic configuration with the highest probability in the $N = 8$ sector
has the high spin ($S = 1$) within the Ni $e_{g}$ state.
Therefore, Hund's physics is still noticeable even for a large $U$ value.

Figure~\ref{fig:optics_U} shows the computed optical conductivity of LaNiO$_{2}$
for some different $U$ values.
Increasing $U$ just reduces the overall magnitude of the optical conductivity
since $U$ suppresses the charge fluctuation.
It is a general characteristic of strongly correlated materials
which shared among many classes of systems,
such as heavy fermions, Mott materials, and Hund's metals.
However, increasing $U$ does not affect
all the interesting spectral weight transfer effects
as a function of temperature and doping which are the signatures of the Hund's metal.
Hence the Hund's picture is still robust even for a large $U$ value.
Note that the $K_{\text{DMFT}}/K_{\text{DFT}} \sim 0.25$
at $\Omega_{c}$ for $U$ = 10 eV.
A well-known Hund’s metal system of BaFe$_{2}$As$_{2}$ has
the kinetic energy ratio with a range from $\sim0.25$ to $\sim0.37$
\cite{Qazilbash2009}.
Hence the $K_{\text{DMFT}}/K_{\text{DFT}}$ value of $\sim0.25$
is still in a regime of the Hund’s system.

Our case for the Hund's metal type of correlations is based on
a large number of additional observations
besides the ratio of the optical spectral weight
(i.e. $K_{\text{DMFT}}/K_{\text{DFT}}$).
We make our case for Hundness on the following additional considerations:
the evolution of the optical spectral weight $K_{\text{DMFT}}$
with (1) temperature and (2) doping,
(3) the DMFT valence histogram with enhanced high spin occupation,
and (4) the strong dependence of the coherence scale with Hund's coupling $J_{H}$.

In addition, we found an orbital differentiation with a temperature dependence of
the effective mass of the correlated $d_{x^2-y^2}$ orbital,
which can be measured in optics.

In fact, it is all these elements pointing towards the same picture together with
the study of an easily measurable observable (optical conductivity) which makes
a compelling case for the Hund's picture of the infinite-layer nickelates.

\section{Comparison with C\lowercase{a}C\lowercase{u}O$_{2}$}
Figure~\ref{fig:histogram_CaCuO2}  shows the DMFT valence histogram for an isostructural CaCuO$_{2}$ material for some different $U$ values.
As shown in Fig.~\ref{fig:histogram_CaCuO2},
Hund’s physics is not notable in CaCuO$_{2}$
since atomic probabilities of $N = 8$ configurations are almost negligible.
Instead, it supports Mott physics.
We would like to note that Hund’s physics is not noticeable in the cuprate
but is noticeable in the infinite-layer nickelate as shown in Fig.~\ref{fig:histogram_U}.

\section{Optical conductivity of a prototypical Mott system of V$_{2}$O$_{3}$}

Figure~\ref{fig:V2O3}(a) shows the optical conductivity of the paramagnetic metallic phase of V$_{2}$O$_{3}$ computed with DFT+DMFT, which is adopted from Ref.~\cite{Deng2014}.
From the data, we compute the integrated optical spectral weight
$K = \int_{0}^{\Omega}\sigma_{1}(\omega)d\omega$ and provide it as a function of integration cutoff value $\Omega$ in Fig.~\ref{fig:V2O3}(b).
Since the electronic kinetic energy is proportional to the integrated optical spectral weight $K$,
Fig.~\ref{fig:V2O3}(b) demonstrates that the kinetic energy decreases upon heating.
This reflects the Mott behavior that the kinetic energy is reduced
as an insulating state is approached at higher temperatures.
This optical response upon heating is opposite to the case of LaNiO$_{2}$,
where the kinetic energy increases upon heating as shown in the main text.
Based on the optical response, the infinite-layer LaNiO$_{2}$ is, therefore, far from a Mott system.

\section{Low-energy physical quantities}
\label{sec:tau}

\begin{table}[b]
\centering
\caption{Quadratic fitting ($aT^2+b$) in the quasiparticle scattering rate $1/\tau_{\text{qp}}$ provided in Fig.~\ref{fig:orbital}(b). The coefficients $a$ and $b$ are provided for each Ni orbital.}
\begin{tabular}{l|| p{1in} | p{1in} }
\hline
\hline
 & $a$ (eV/K$^2$) & $b$ (eV) \\
\hline
$d_{x^2-y^2}$   & $2.029 \times 10^{-7}$ & $6.124 \times 10^{-5}$ \\
$d_{z^2}$       & $4.563 \times 10^{-9}$ & $2.532 \times 10^{-4}$ \\
$d_{xz}/d_{yz}$ & $1.753 \times 10^{-9}$ & $2.045 \times 10^{-4}$ \\
$d_{xy}$        & $2.393 \times 10^{-9}$ & $2.117 \times 10^{-4}$ \\
\hline
\hline
\end{tabular}
\label{tab:fitting}
\end{table}

Figure~\ref{fig:Sig} shows the imaginary part of self-energy for Ni $d_{x^2-y^2}$ on the real axis. The modified Gaussian method~\cite{Haule2010-prb} is adopted for analytical continuation.
Upon cooling, $\text{Im}\Sigma(0)$ approaches to zero and $\text{Im}\Sigma(\omega)$ exhibits a quadratic behavior at low frequencies.
Hence, LaNiO$_{2}$ clearly shows the Fermi liquid behavior.

The quasiparticle weight $Z$ and the quasiparticle scattering rate $1/\tau_{\text{qp}}=-Z\text{Im}\Sigma(i0^{+})$ are depicted in Fig.~\ref{fig:orbital}.
Ni $d_{x^2-y^2}$ has the lowest $Z$ (0.4 $\sim$ 0.5) among the others
($Z \approx$ 0.8),
indicating that Ni $d_{x^2-y^2}$ is the correlated orbital
and the others are almost uncorrelated ones.
$1/\tau_{\text{qp}}$ for all Ni $3d$ orbitals exhibit a quadratic behavior in temperature, thereby presenting the Fermi liquid behavior.
In the case of Ni $d_{x^2-y^2}$, deviation from the quadratic behavior in $1/\tau_{\text{qp}}$ is recognized around $T \sim 600$ K.
The coherent temperature is $T_{coh} \sim 450 K$ as demonstrated in Fig.~\ref{fig:orbital}(c),
hence deviation from the Fermi liquid behavior is apparent above the coherent temperature.
It is noteworthy that the only Ni $d_{x^2-y^2}$ orbital shows coherence-incoherence crossover and the others are still coherent even at high temperature.
The orbital-differentiated coherence-incoherence crossover is one of the hallmark of a Hund's metal.

Table~\ref{tab:fitting} provides coefficients of the quadratic fitting
($aT^2+b$) on $1/\tau_{\text{qp}}$ used in Fig.~\ref{fig:orbital}(b).
For Ni $d_{x^2-y^2}$, $b$ is relatively small and $\Gamma/k_{B}T$ depicted in Fig.~\ref{fig:orbital}(c) follows $aT+b/T \approx aT$, that is a linear behavior.
On the other hand, the others have relatively large $b$,
so that they exhibit $b/T$ at low temperature, but $aT$ at high temperature
as demonstrated in Fig.~\ref{fig:orbital}(c).

\section{Electronic structure of the infinite-layer nickelate upon Strontium doping}

Figure~\ref{fig:dmft-FS} shows the 3-dimensional Fermi surface of the infinite-layer La$_{1-x}$Sr$_{x}$NiO$_{2}$ for several doping concentrations $x$ computed with DFT+DMFT.
$\text{Im}\Sigma(\omega)$ is set to be zero in Fig.~\ref{fig:dmft-FS},
which leads to obtain the quasiparticle Fermi surface.
At $x=0.1$, the small electron pocket at $\Gamma$ disappears.
Upon further doping, another electron pocket at $A$ disappears at $x=0.5$.
Therefore, two distinct Lifshitz transitions are realized from $x=0.0$ to $x=0.5$.

The low-energy carrier number, $n_{e}$, is defined by the volume of the Fermi surface.
The dependence of the Fermi surface with doping
is provided in Fig.~\ref{fig:dmft-FS}.
The size of the Ni $d_{x^2-y^2}$ Fermi surface shrinks upon doping and its $n_{e}$ decreases accordingly. It reduces the electronic kinetic energy as shown in Fig. 3 in the main text.
It is another definite evidence of that the infinite-layer La$_{1-x}$Sr$_{x}$NiO$_{2}$ is far from a Mott system, where doping increases the electronic kinetic energy.

\section{Hund's rule correlation in the infinite-layer nickelate}

In order to gain more insight into the role of Hund's rule correlation in the infinite-layer LaNiO$_{2}$, we provide the DMFT valence histogram as a function of Hund's coupling $J_{H}$ as depicted in Fig.~\ref{fig:histogram}(a).
Note that DMFT simulations with $J_{H}$ = 0.3 and 0 eV along with the exact double counting scheme give different occupation numbers of Ni $3d$ orbitals, $n_{d}$ = 8.29, and 8.17, respectively. Recall that $n_{d}$ = 8.58 for $J_{H}$ = 1.0 eV.
Therefore, in order to balance the $n_{d}$ for $J_{H}$ = 1.0 eV and that for other $J_{H}$ values, we fix the double-counting energy obtained in the $J_{H}$ = 1.0 eV simulation and perform DMFT calculations with other $J_{H}$ values.
It helps to obtain the similar $n_{d}$.
As demonstrated in Fig.~\ref{fig:histogram}(a), the most probable atomic configuration in a $N$ = 8 sector is the spin-triplet state ($S$ = 1) in Ni $e_{g}$ orbitals and its probability decreases as $J_{H}$ becomes smaller.
It clearly shows Hund's rule correlation.
Hund's rule correlation is clearly exhibited in Figs.~\ref{fig:histogram}(b) and (c), where the quasiparticle weight $Z$
and the electronic coherence scale $-\text{Im}\Sigma(i0^{+})$
change significantly as a function of Hund's coupling $J_{H}$.
Figure~\ref{fig:histogram}(c) particularly shows that Hund's coupling $J_{H}$ increases $-\text{Im}\Sigma(i0^{+})$, thereby reducing the coherence scale drastically.
It leads to lower the coherent temperature $T_{coh}$.
In Hund's metal systems such as ruthenates and iron pnictides and chalcogenides,
the same feature of the reduction of the coherence scale due to $J_{H}$ is observed.
Therefore, the infinite-layer LaNiO$_{2}$ is definitely a Hund's metal.

\section{Coherence-Incoherence Crossover}
Figure~\ref{fig:coherence}(a) shows $\Gamma/k_{B}T$ as a function of $T$,
where $\Gamma$ is the quasiparticle scattering rate with
$\Gamma = -Z\text{Im}\Sigma(i0^{+})$.
The Fermi-liquid behavior of $\Gamma \propto T^{2}$ is identified up to $\sim400$ K.
The coherence temperature $T_{coh}$ is defined by $\Gamma/k_{B}T = 1$
and it is estimated as $T_{coh} \sim 450$ K.
The deviation from the $T^{2}$ behavior is visible above the coherence temperature.
The same behavior could be found in $1/\tau_{\text{tr}}^{*}$ for Ni $d_{x^2-y^2}$
as shown in Fig. 2(e) of the main text.
Note that the coherence scale (or $T_{coh}$)
is severely diminished by Hund's coupling $J_{H}$
as demonstrated in Fig.~\ref{fig:histogram}(c).
The coherence-incoherence crossover is already realized in other Hund's metal systems
such as ruthenates~\cite{Mravlje2011,Mravlje2016} and iron pnictides~\cite{Haule2009-njp,Hardy2013,Miao2016} and chalcogenides~\cite{Liu2015},
where Hund's coupling $J_{H}$ drastically reduces the coherence scale.

The coherence temperature $T_{coh}$ manifests itself also in the $T$-dependent Fermi surface (FS),
displayed in Fig.~\ref{fig:coherence}(b).
Below $T_{coh}$, quasiparticles are well-defined, thereby providing apparent FSs.
Above $T_{coh}$, Ni $3d$ orbitals except for $d_{x^2-y^2}$ are still coherent even at very high temperature, however the Ni $d_{x^2-y^2}$ gets incoherent.
$T_{coh}$ is identified in $(\omega_{p}^{*})^2$ as well presented in Fig. 2(d) of the main text,
where $(\omega_{p}^{*})^2$ for Ni $d_{x^2-y^2}$ increases linearly upon heating below $T_{coh}$ and then shows the saturation behavior above $T_{coh}$.
$(\omega_{p}^{*})^2$ for the uncorrelated hybridized band does not show temperature dependence.
The orbital-differentiated coherence-incoherence crossover is also
observed in other Hund's metal systems~\cite{Mravlje2011,Miao2016}.

\section{Two-band model}
\label{sec:two-band}

We investigate the effective low-energy Hamiltonian of the infinite-layer LaNiO$_{2}$
based on the Wannier interpretation~\cite{wannier90-1,wannier90-2}.
Since band characters that cross the Fermi level are the correlated Ni $d_{x^2-y^2}$ and the uncorrelated hybridized band including Ni $d_{z^2}$, we choose two Wannier bases of Ni $d_{x^2-y^2}$ and $d_{z^2}$ orbitals.
A small energy window of (-1.5 $\sim$ 2.5 eV) is set for Wannier construction.
Note that a large energy window covering the whole Ni $3d$ bands leads to
a two-band model proposed by Wan \emph{et al.} \cite{Wan2020-arXiv}
and Sakakibara \emph{et al.} \cite{Sakakibara20-prl-suppl},
where it does not fully describe the DFT band near the Fermi level.
During the process of Wannier minimization,
we found that a center of the Wannier wave-function of Ni $d_{z^2}$ is shifted toward a La site and the spread of the Wannier wave-function develops gradually due to a significant hybridization with La $d_{z^2}$ orbital.
Hence, in order to keep an atom centered Ni $d_{z^2}$ orbital,
we performed an one-shot Wannier minimization computation and the results are presented in Fig.~\ref{fig:wannier}.
As shown in Fig.~\ref{fig:wannier}, two Wannier bands explains the DFT bands near $E_{\text{F}}$, thereby suggesting that the effective low-energy Hamiltonian is described with atom centered Ni $e_{g}$ orbitals.
Recall that orbital characters of the hybridized band crossing $E_{\text{F}}$
around $A$-point are Ni $d_{xz}/d_{yz}$ and La $d_{xy}$ (see Fig.~\ref{fig:dft-band}).
The corresponding DFT charge density
has symmetry close to the Ni $d_{z^2}$.
That is the reason why the two-band model with Ni $e_{g}$ orbitals reproduces the DFT band with different orbital characters.

It is noteworthy that alternative low-energy Wannier constructions have been reported.
Aside from the two-band model proposed by Wan \emph{et al.} \cite{Wan2020-arXiv}
and Sakakibara \emph{et al.} \cite{Sakakibara20-prl-suppl},
there are a three-band model composing of Ni $d_{x^2-y^2}$, rare-earth $d_{z^2}$, and interstitial $s$ orbitals~\cite{Nomura2019-prb-suppl},
a two-band model with Ni $d_{x^2-y^2}$ and rare-earth $d_{z^2}$ orbitals~\cite{Lee2020-nm-suppl},
and a two-band model with Ni $d_{x^2-y^2}$ and an axial orbital~\cite{Adhikary2020-prb-suppl}.
In that sense, our two-band model with atom centered Ni $e_{g}$ orbitals is quite unique.

\section{Dynamical Crystal Field Splitting}

Figure~\ref{fig:Eimp} shows the frequency dependent (dynamical)
crystal field splitting of Ni $e_g$ orbitals in LaNiO$_{2}$.
It is computed from the difference in
the dynamical crystal field energy level, $E_{\text{CF}}(i\omega)$,
between the $e_g$ orbitals.
$E_{\text{CF}}(i\omega)$ could be expressed as
\begin{equation}
E_{\text{CF}}(i\omega)
= E_{\text{imp}} + \big( \text{Re}[\Sigma_{\text{imp}}(i\omega)]
- E_{dc} \big) + \text{Re}[\Delta(i\omega)],
\end{equation}
where $E_{\text{imp}}$, $\Sigma_{\text{imp}}(i\omega)$, $\Delta(i\omega)$,
and $E_{dc}$ are the impurity level, the impurity self-energy,
the hybridization function, and the double counting energy, respectively,
which could be obtained from fully converged DFT+DMFT calculations.

At the zero frequency, a large crystal field splitting of 2.90 eV is realized.
It is due to the absence of apical oxygens.
This large crystal field splitting becomes smaller as frequency becomes larger.
Notice that there is the competition between the crystal field splitting
and Hund’s coupling within the Ni $e_g$ orbitals.
Hence, Hund’s physics becomes more favorable at higher frequencies.
When the crystal field splitting becomes small
and comparable to the strength of Hund's coupling $J_{H} \sim$ 1 eV,
a crossover from Mott to Hund region happens.
This type of Hund's picture is also well captured in the DMFT valence histogram
as shown in Fig.~\ref{fig:histogram_U}.
Therefore, Hund's correlation is hidden at low energies
but noticeable at intermediate energies.
It is different from Hund's metals realized in iron pnictides, chalcogenides,
and ruthenates, where Hund's correlation is prominent at low energies.

Figure~\ref{fig:CaCuO2-Eimp} shows the dynamical crystal field splitting
of Cu $e_g$ orbitals in the isostructural CaCuO$_{2}$ material.
It shows the sufficiently larger crystal field splitting than LaNiO$_{2}$ so that,
unlike LaNiO$_{2}$, there is no crossover from Mott to Hund region.
This makes Cu $d_{x^2-y^2}$ the only active orbital over the large energy window,
validating a one band description.
Therefore, Hund's physics is not noticeable in the cuprate
while Mott physics is evident therein.
This picture is clearly demonstrated in Fig.~\ref{fig:histogram_CaCuO2} as well.

\section{Hall coefficient as a function of doping and temperature}

\begin{figure*}
\centering
\includegraphics[width=0.90\textwidth]{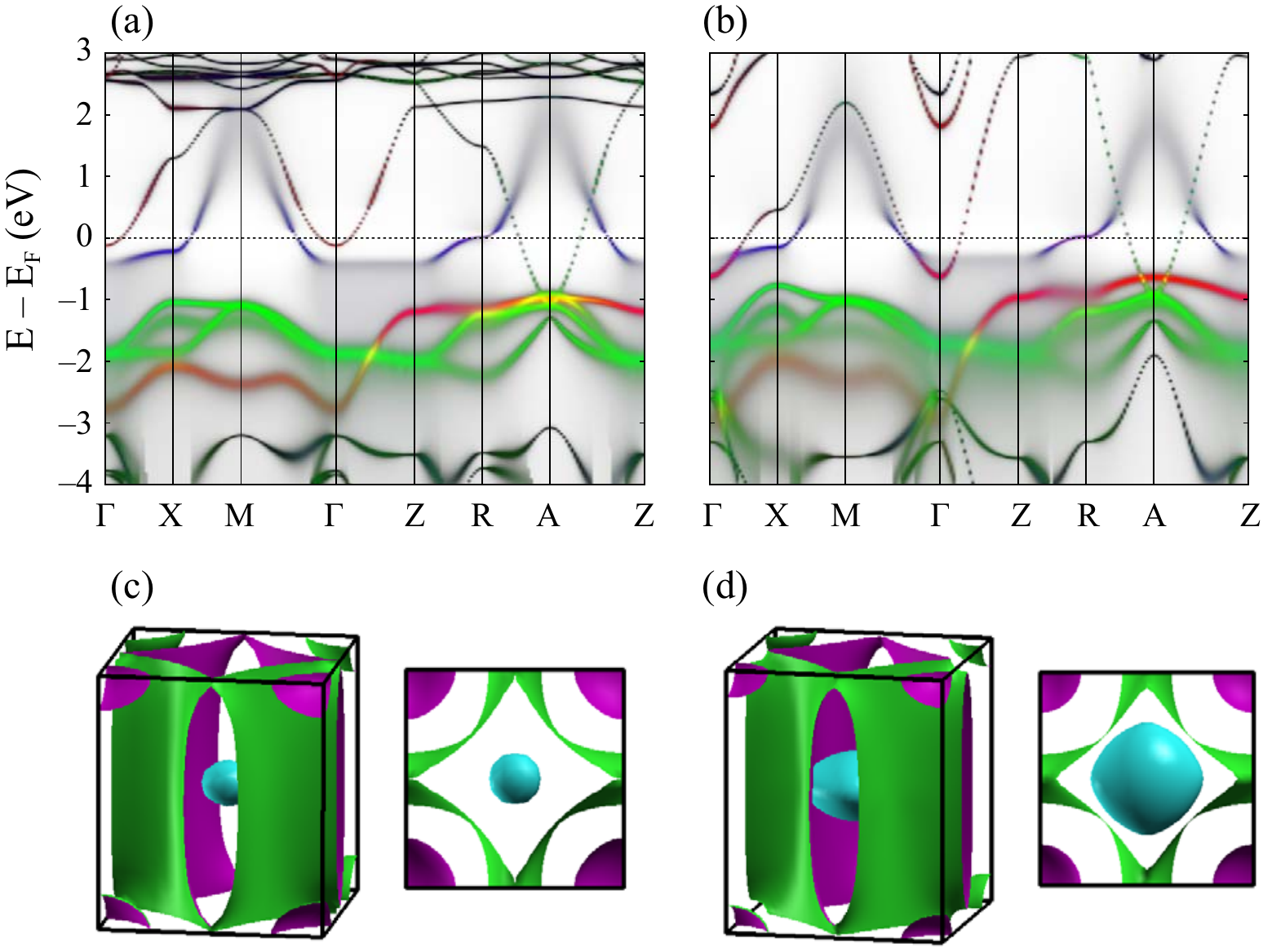}
\caption{Electronic structures computed within DFT+DMFT at $T$ = 116 K.
$k$-resolved spectral functions for (a) LaNiO$_{2}$ and (b) NdNiO$_{2}$.
Note that Nd $4f$ electrons are treated as core electrons in the calculations.
Ni $d_{z^2}$, $d_{x^2-y^2}$, and $t_{2g}$ orbital characters are presented by red, blue, and green, respectively, in (a) and (b).
Fermi surfaces of (c) LaNiO$_{2}$ and (d) NdNiO$_{2}$.
Top view is provided for comparison.
}
\label{fig:ndnio2-dmft}
\end{figure*}

\begin{figure*}
\centering
\includegraphics[width=0.8\textwidth]{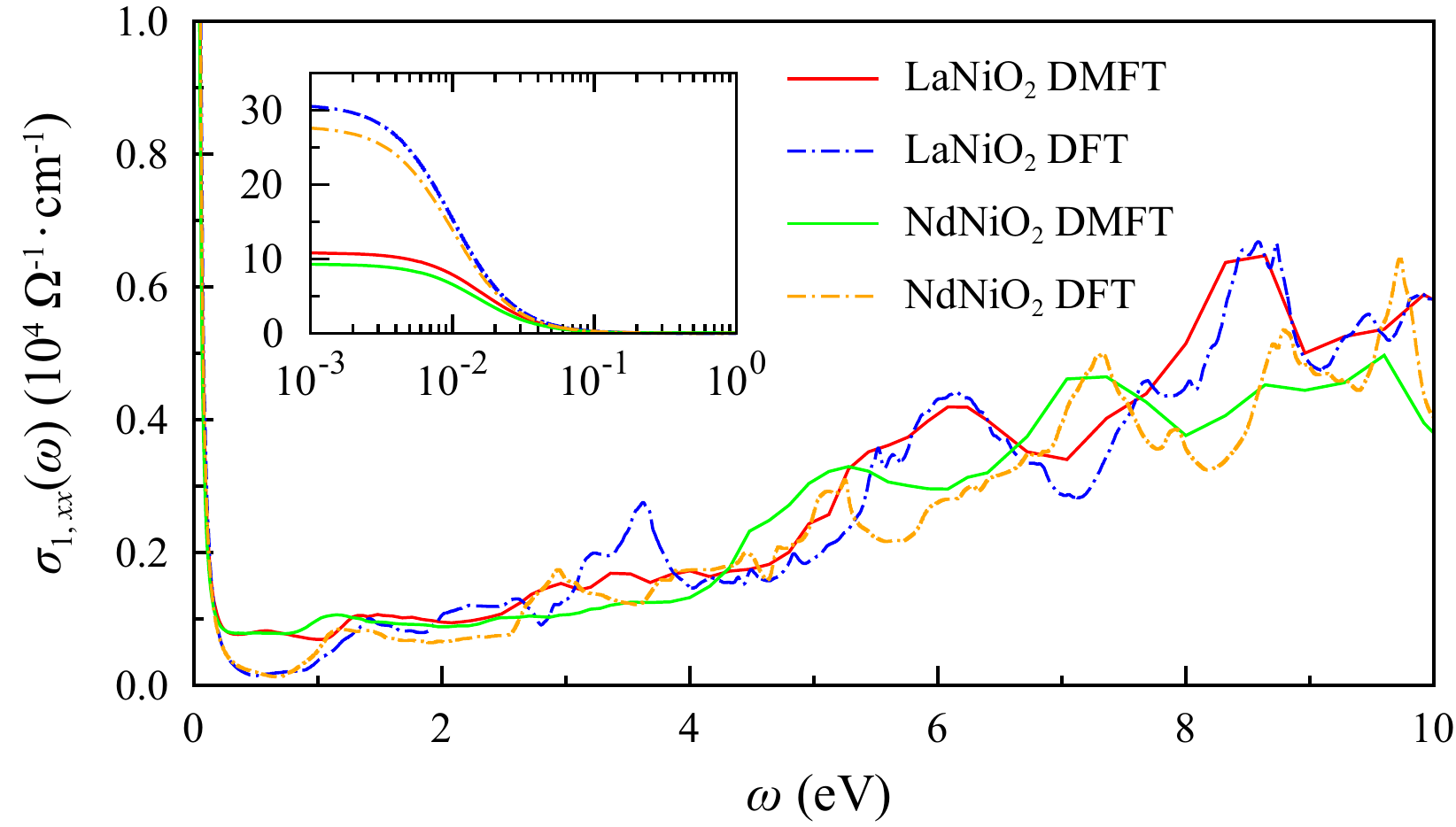}
\caption{Optical conductivities of LaNiO$_{2}$ and NdNiO$_{2}$ in a broad $\omega$ frequency range computed within both DFT and DFT+DMFT (at $T$ = 116 K) methods.
Note that Nd $4f$ electrons are treated as core electrons in the calculations.
The inset is provided to magnify the Drude weight.
}
\label{fig:ndnio2-optics}
\end{figure*}

\begin{figure*}
\centering
\includegraphics[width=0.9\textwidth]{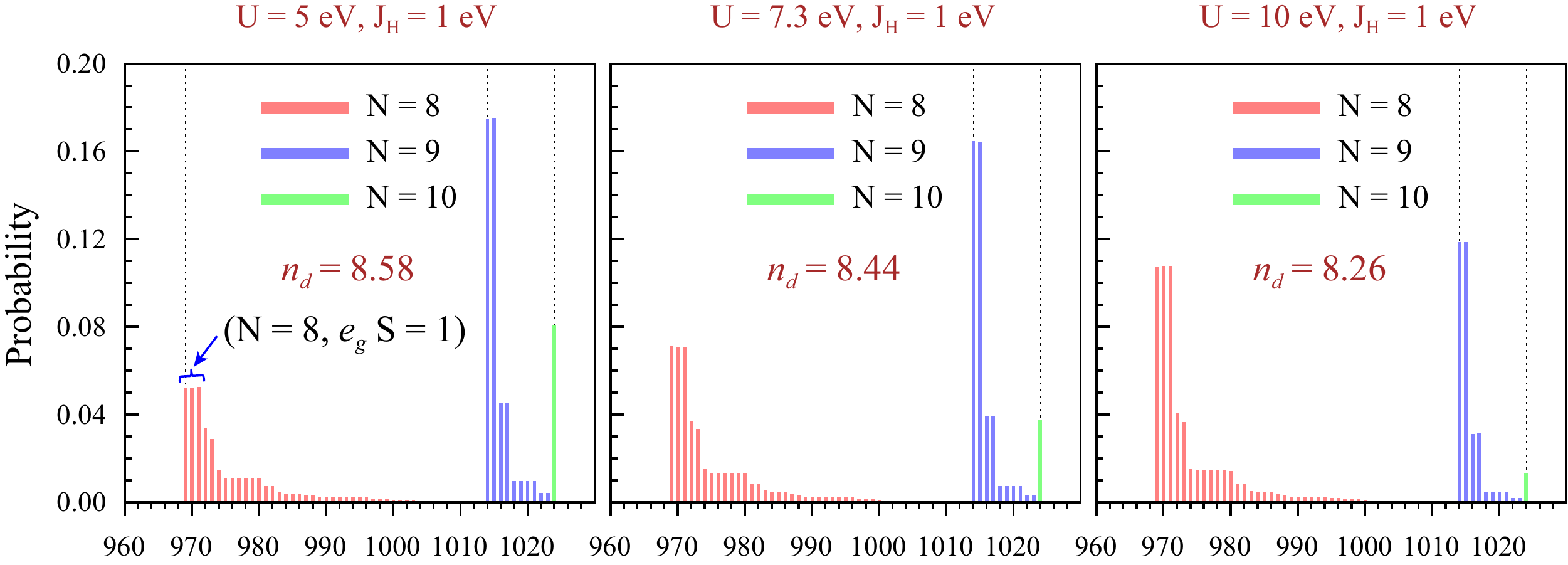}
\caption{DMFT valence histogram of the Ni-$3d$ shell in LaNiO$_{2}$ for different Coulomb $U$ values.
The 1024 possible atomic configurations are sorted by number of $3d$ electrons of the individual configuration.
An atomic configuration with the highest probability in the $N = 8$ sector has the high spin ($S$ = 1)
within the $e_{g}$ state.
Hence, Hund’s physics is still noticeable even for a large $U$ value.
}
\label{fig:histogram_U}
\end{figure*}

\begin{figure*}
\centering
\includegraphics[width=0.8\textwidth]{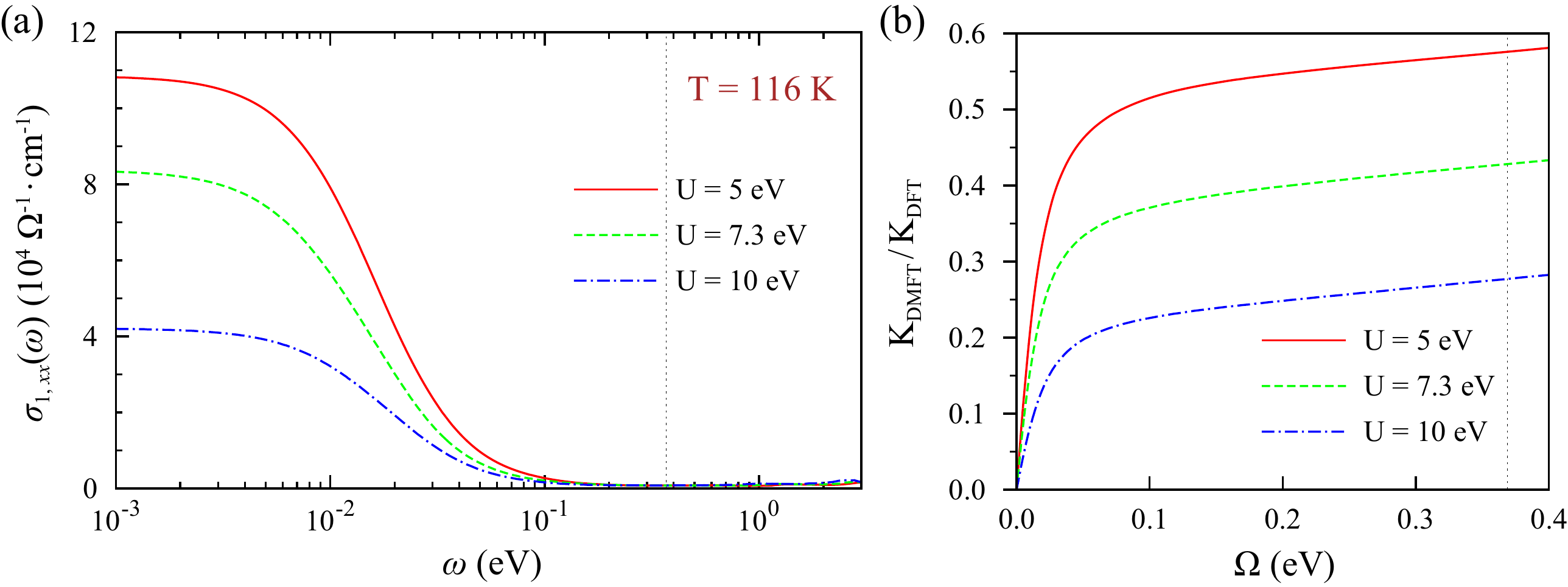}
\caption{Optical conductivity of LaNiO$_{2}$ computed within the DFT+DMFT method
(T = 116 K is selected and $J_H$ = 1 eV is fixed in the calculations).
(a) Optical conductivity for different Coulomb $U$ values.
The $\omega$-axis is presented in a logarithmic scale.
(b) The kinetic energy ratio $K_{\text{DMFT}}/K_{\text{DFT}}$
as a function of integration cutoff value $\Omega$ provided for different $U$ values.
The vertical dotted line is the kinetic energy intergration cutoff
$\Omega_{c}$ = 0.369 eV
chosen to exclude a contribution from interband transitions.
}
\label{fig:optics_U}
\end{figure*}

\begin{figure*}
\centering
\includegraphics[width=0.9\textwidth]{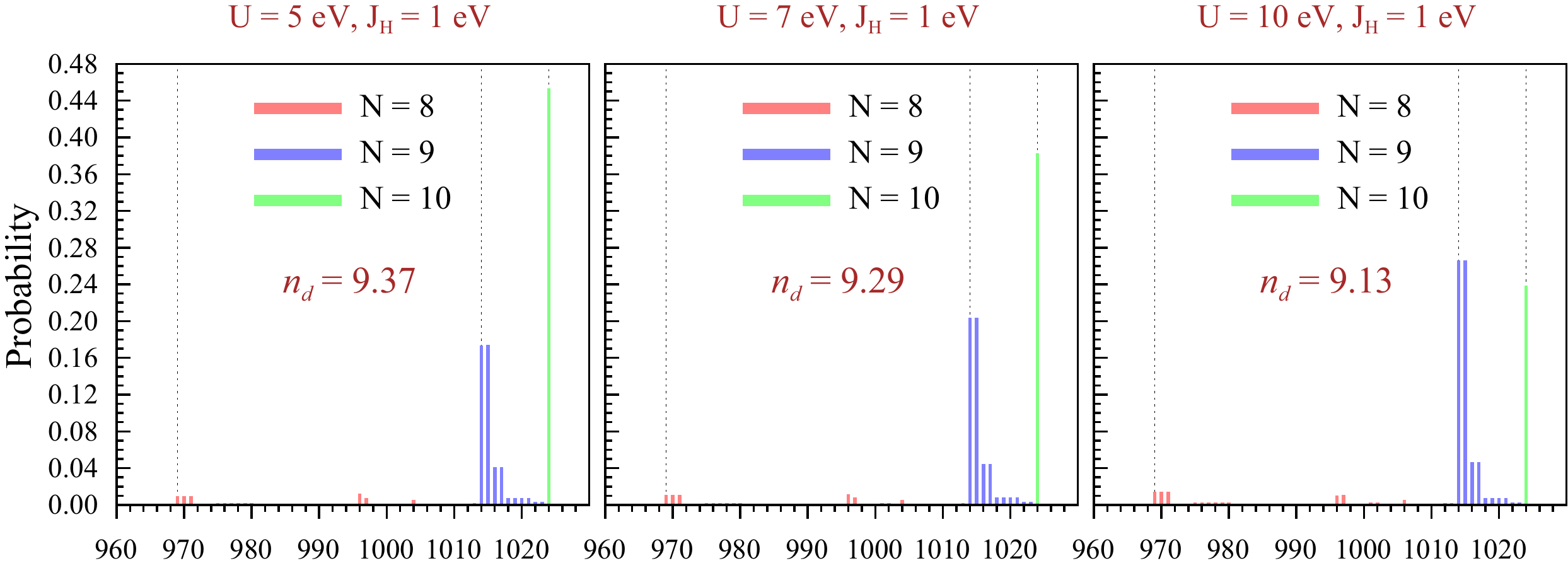}
\caption{DMFT valence histogram of the Cu-$3d$ shell in CaCuO$_{2}$ for different Coulomb $U$ values.
The 1024 possible atomic configurations are sorted by number of $3d$ electrons of the individual configuration.
Hund's physics is not noticeable in the cuprate while Mott physics is evident therein.
}
\label{fig:histogram_CaCuO2}
\end{figure*}

\begin{figure*}
\centering
\includegraphics[width=0.8\textwidth]{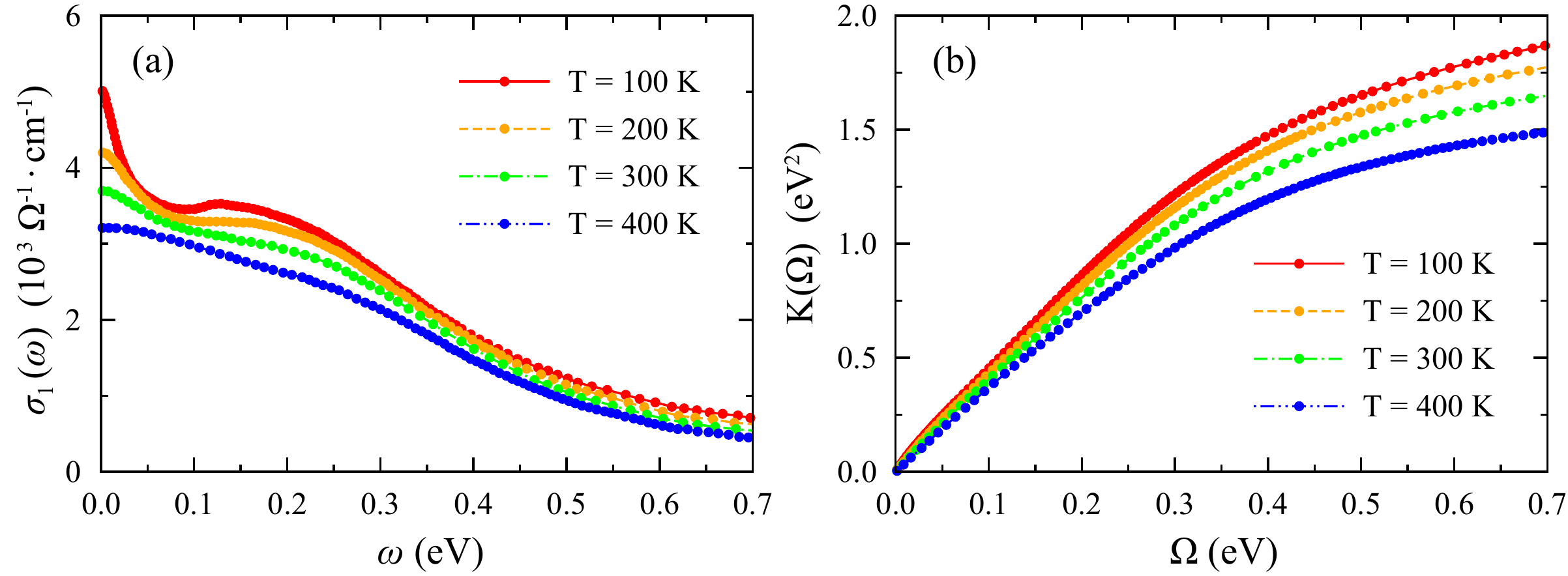}
\caption{
(a) The optical conductivity of the paramagnetic metallic phase of V$_{2}$O$_{3}$ computed within the DFT+DMFT method. The data is adopted from Ref.~\cite{Deng2014}.
(b) The integrated optical spectral weight $K(\Omega) = \int_{0}^{\Omega}\sigma_{1}(\omega)d\omega$ as a function of cutoff frequency $\Omega$.
It decreases upon heating, which is a key characteristic of a Mott system.
}
\label{fig:V2O3}
\end{figure*}

\begin{figure*}
\centering
\includegraphics[width=0.5\textwidth]{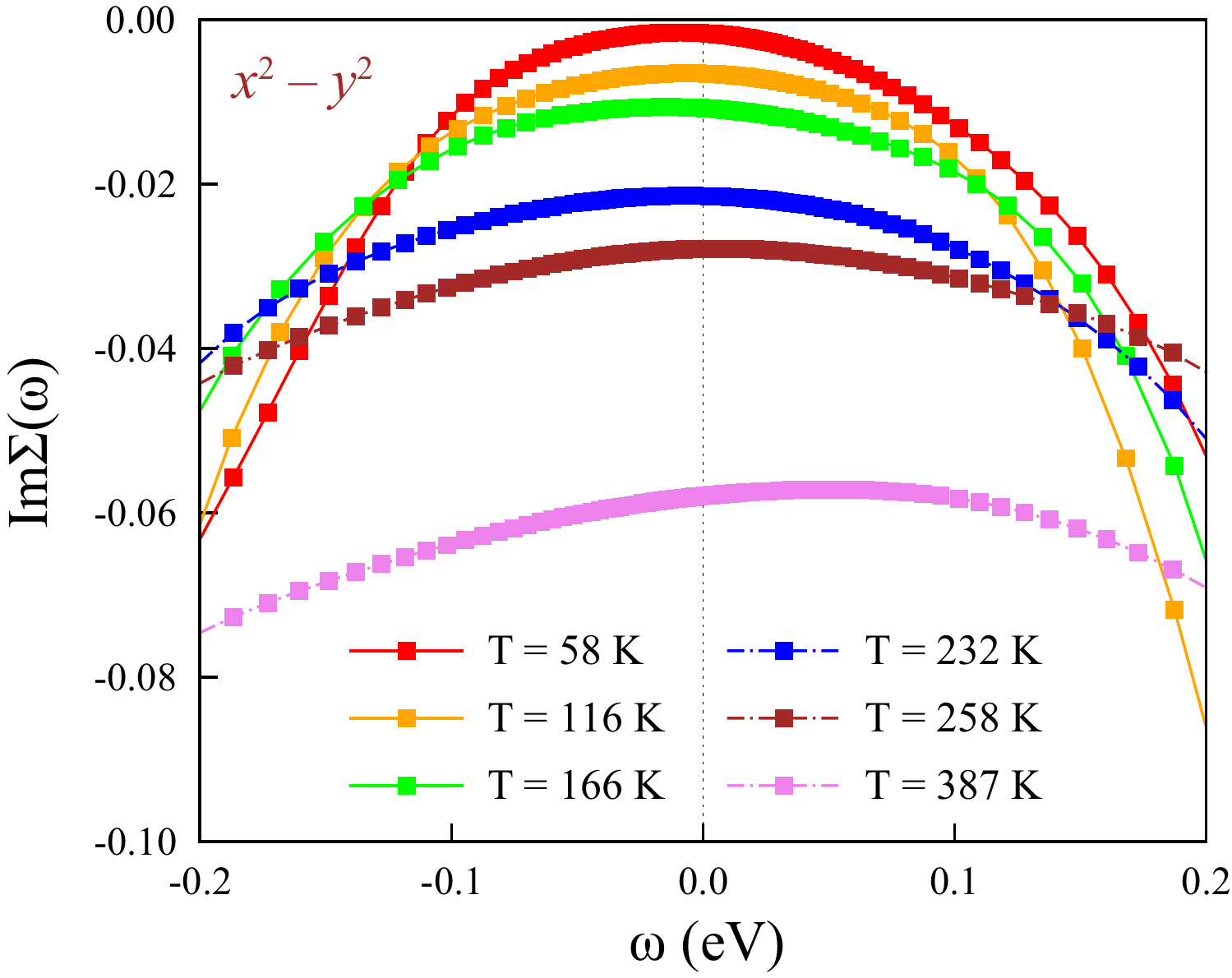}
\caption{DMFT self-energy in the infinite-layer LaNiO$_{2}$.
The imaginary part of self-energy for Ni $d_{x^2-y^2}$ orbital is presented on the real axis for several temperatures.
}
\label{fig:Sig}
\end{figure*}

\begin{figure*}
\centering
\includegraphics[width=0.95\textwidth]{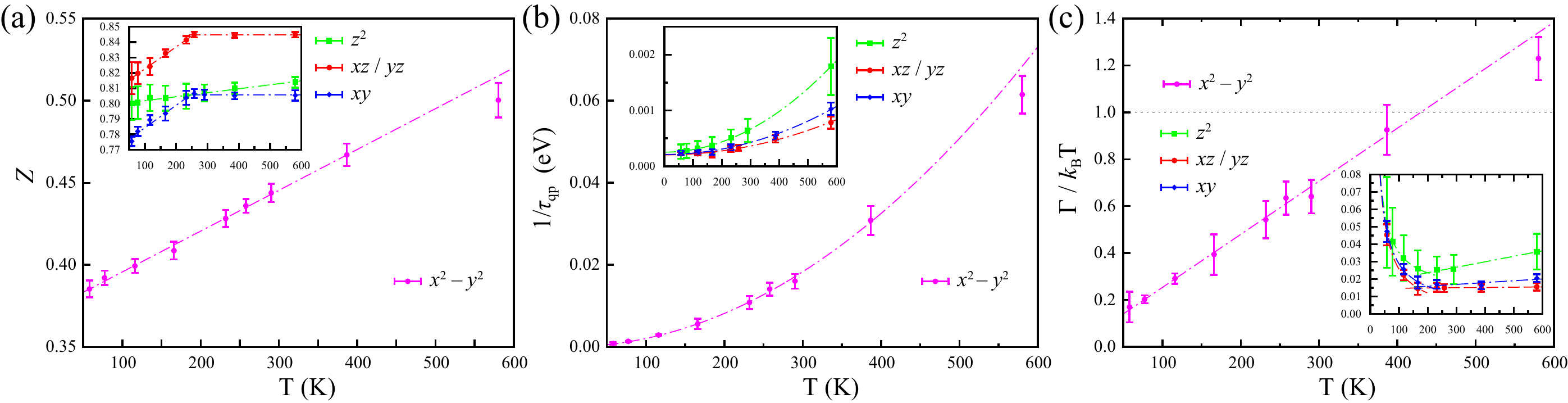}
\caption{Low-energy physical quantities for Ni $3d$ orbitals in the infinite-layer LaNiO$_{2}$.
(a) Quasiparticle weight $Z$ as a function of temperature.
(b) Quasiparticle scattering rate $1/\tau_{\text{qp}}=-Z\text{Im}\Sigma(i0^{+})$ as a function of temperature.
(c) Quasiparticle scattering rate $\Gamma=1/\tau_{\text{qp}}$ divided by $k_{B}T$ as a function of temperature.
Error bars presented in (a), (b), and (c) originate from the statistical errors in CTQMC simulations.
The dash-dotted lines in (a) and (b) are guides for the eye by fitting $Z$ and $1/\tau_{\text{qp}}$ to linear ($aT + b$) and quadratic ($aT^2+b$) functions, respectively.
The dash-dotted lines in (c) are guides for the eye and have a form of $aT+b/T$ where the coefficients $a$ and $b$ are obtained from the fitting in (b).
}
\label{fig:orbital}
\end{figure*}

\begin{figure*}
\centering
\includegraphics[width=0.95\textwidth]{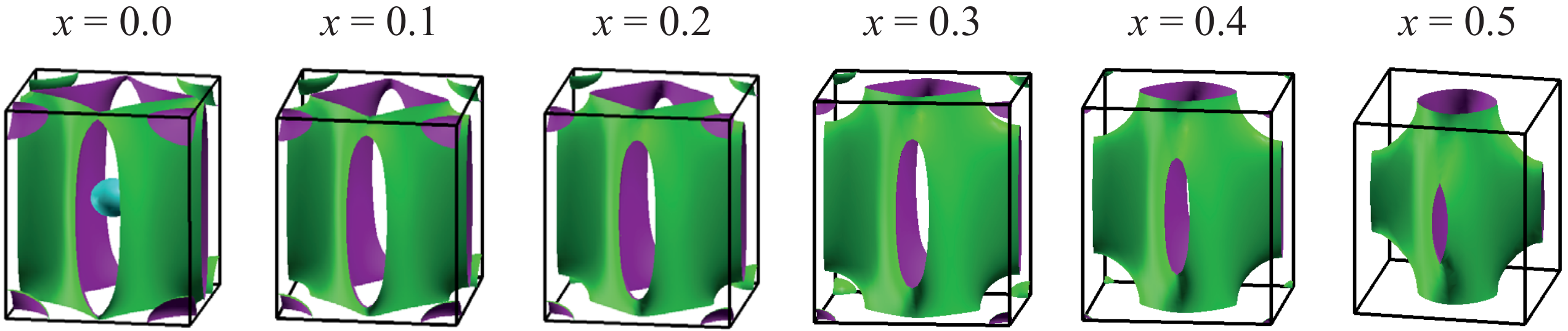}
\caption{DMFT Fermi surfaces of the infinite-layer La$_{1-x}$Sr$_{x}$NiO$_{2}$ as a function of doping ratio $x$.
In this figure, Im$\Sigma(\omega)$ is set to be zero to obtain the quasiparticle DMFT Fermi surfaces.
}
\label{fig:dmft-FS}
\end{figure*}

\begin{figure*}
\centering
\includegraphics[width=0.95\textwidth]{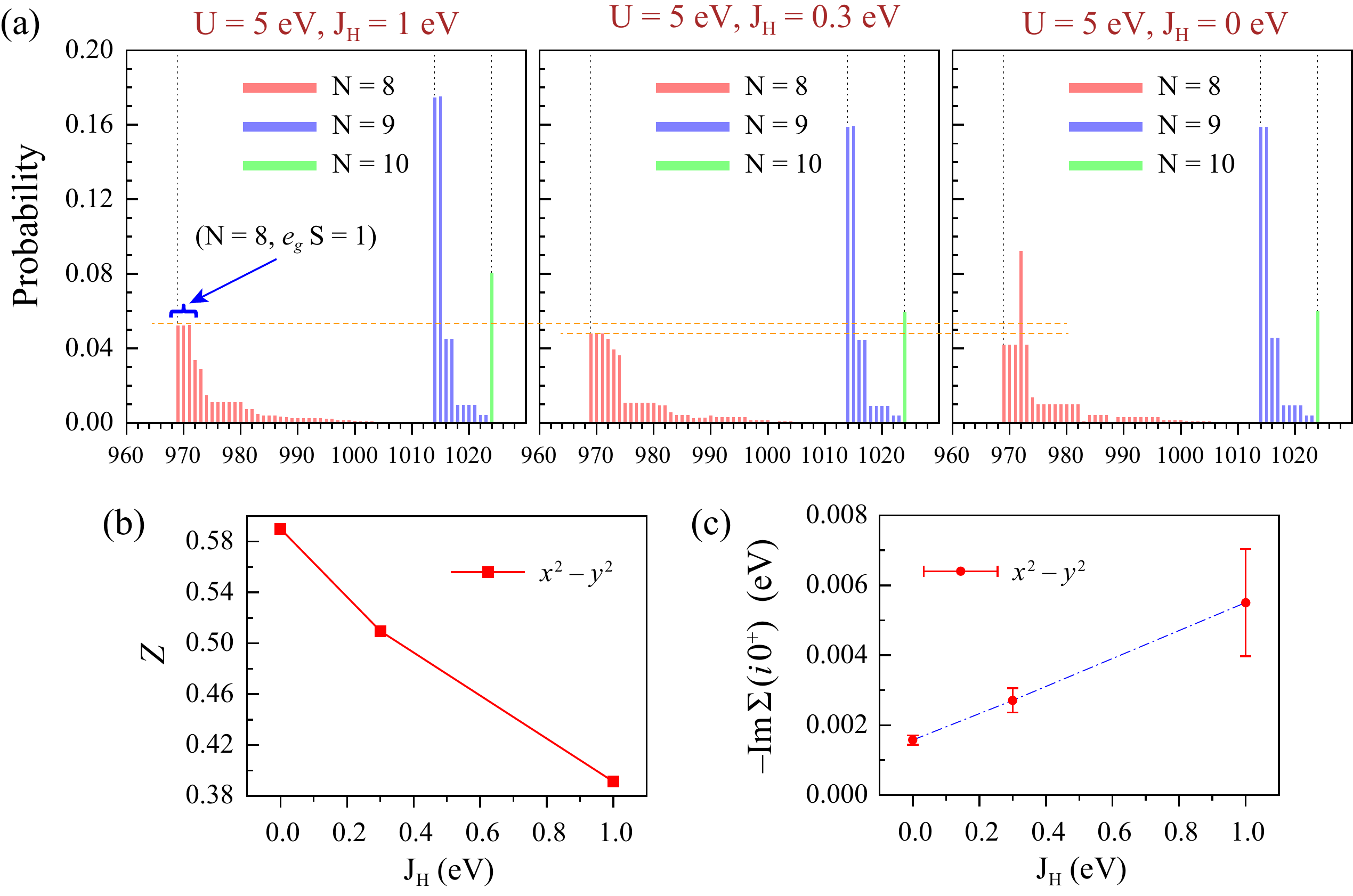}
\caption{Hund's rule correlation in the infinite-layer LaNiO$_{2}$.
(a) The DMFT valence histogram of the Ni-$3d$ shell for LaNiO$_{2}$ is provided for Hund's coupling J$_{H}$ = (left) 1, (middle) 0.3, and (right) 0 eV.
The 1024 possible atomic configurations are sorted by the number of $3d$ electrons of the individual configuration.
The probability of the atomic configuration of (N = 8, $e_{g}$ S = 1), where the spin triplet state is realized in Ni $e_{g}$ orbitals, decreases as $J_{H}$ becomes smaller.
(b) Quasiparticle weight $Z$ as a function of Hund's coupling $J_{H}$.
(c) Electronic coherence scale $-\text{Im}\Sigma(i0^{+})$ as a function of Hund's coupling $J_{H}$.
The dash-dotted line in (c) is a guide for the eye.
Error bars originate from the statistical errors in CTQMC simulations.
}
\label{fig:histogram}
\end{figure*}

\begin{figure*}
\centering
\includegraphics[width=0.95\textwidth]{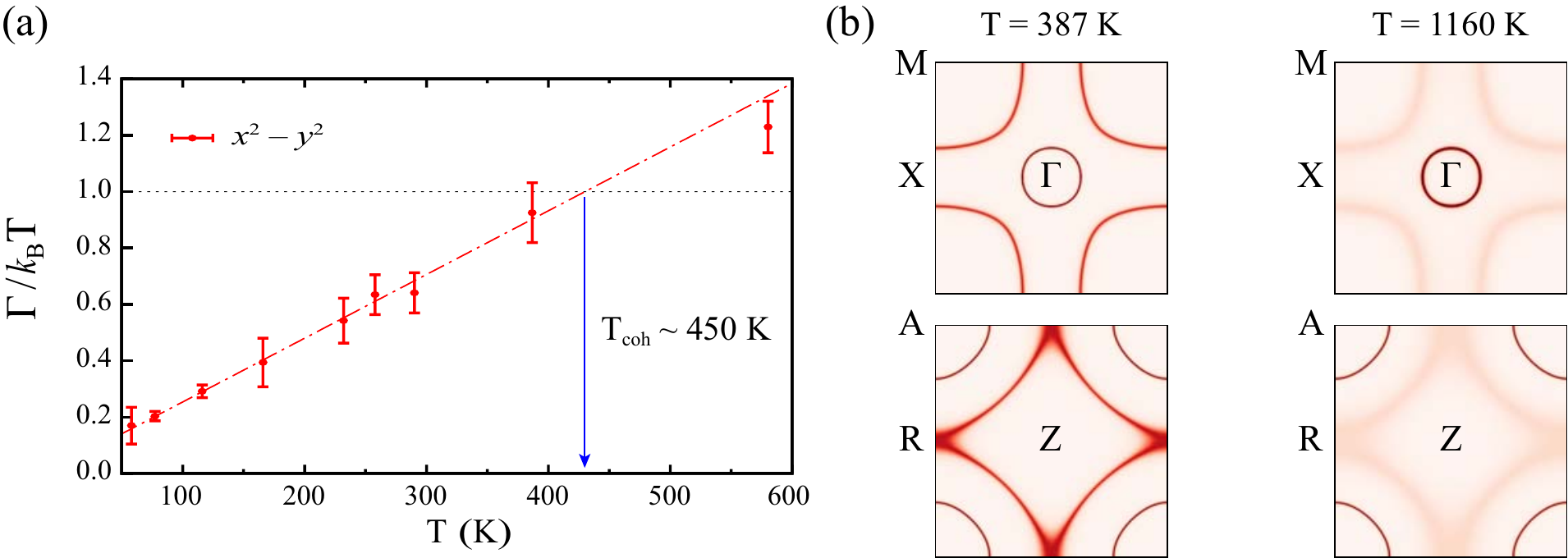}
\caption{
Coherence-incoherence crossover in LaNiO$_{2}$.
(a) Quasiparticle scattering rate $\Gamma = -Z \text{Im}\Sigma(i0^{+})$ of Ni $d_{x^2-y^2}$
divided by $k_{B}T$ as a function of temperature.
Error bars originate from the statistical errors in CTQMC simulations.
The red dash-dotted line is a guide for the eye by fitting $\Gamma/k_{B}T$ to a linear function.
$\Gamma/k_{B}T \approx 1$ at the coherent temperature $T_{coh} \sim 450 K$.
Above $T_{coh}$, $\Gamma/k_{B}T$ shows deviation from linearity,
indicating that a quasiparticle is no longer well-defined.
The coherence-incoherence crossover is clearly shown in (b),
where Fermi surfaces are plotted at temperature below and above $T_{coh}$.
}
\label{fig:coherence}
\end{figure*}

\begin{figure*}
\centering
\includegraphics[width=0.95\textwidth]{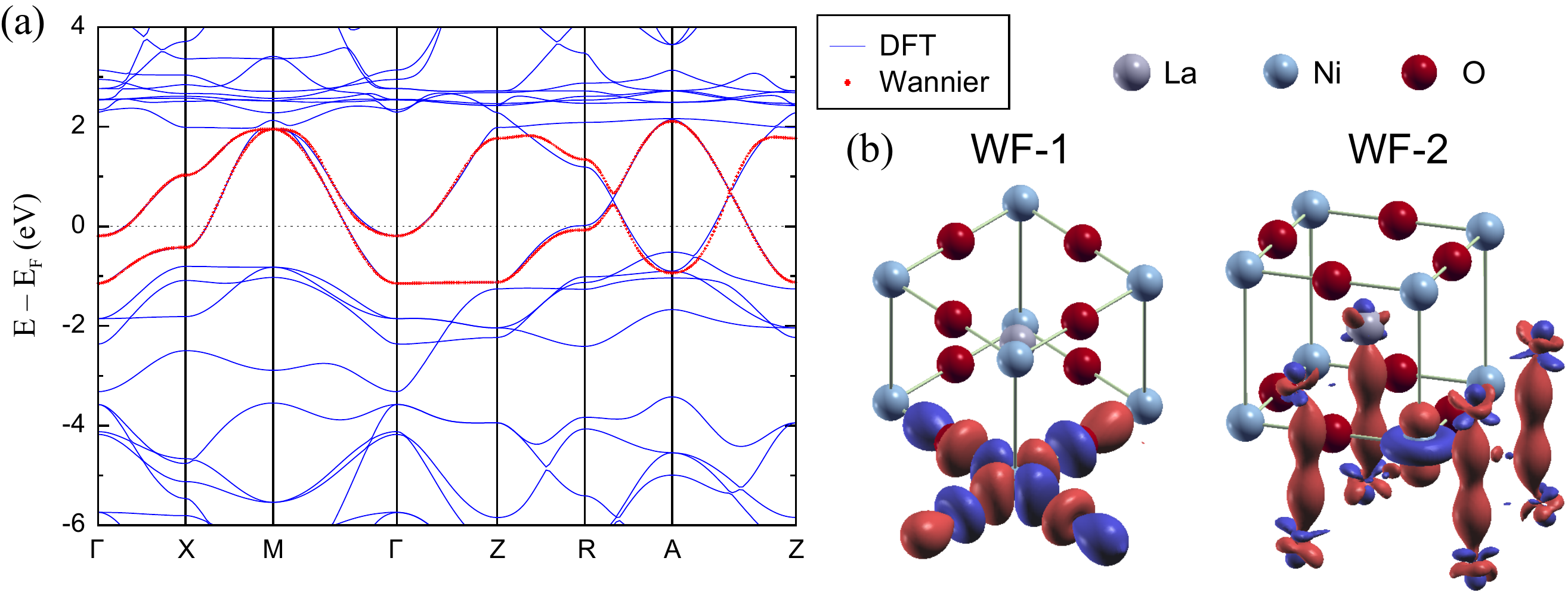}
\caption{Two-band model.
(a) DFT (blue solid line) and Wannier (red dot) band dispersions of LaNiO$_{2}$.
(b) Isosurface plots for (left) Ni-centered $d_{x^2-y^2}$-like and
(right) Ni-centered $d_{z^2}$-like with extended La-centered $d_{z^2}$-like Wannier orbitals.
These two orbitals describe the DFT band dispersion well near the Fermi level in (a).
}
\label{fig:wannier}
\end{figure*}

\begin{figure*}
\centering
\includegraphics[width=0.9\textwidth]{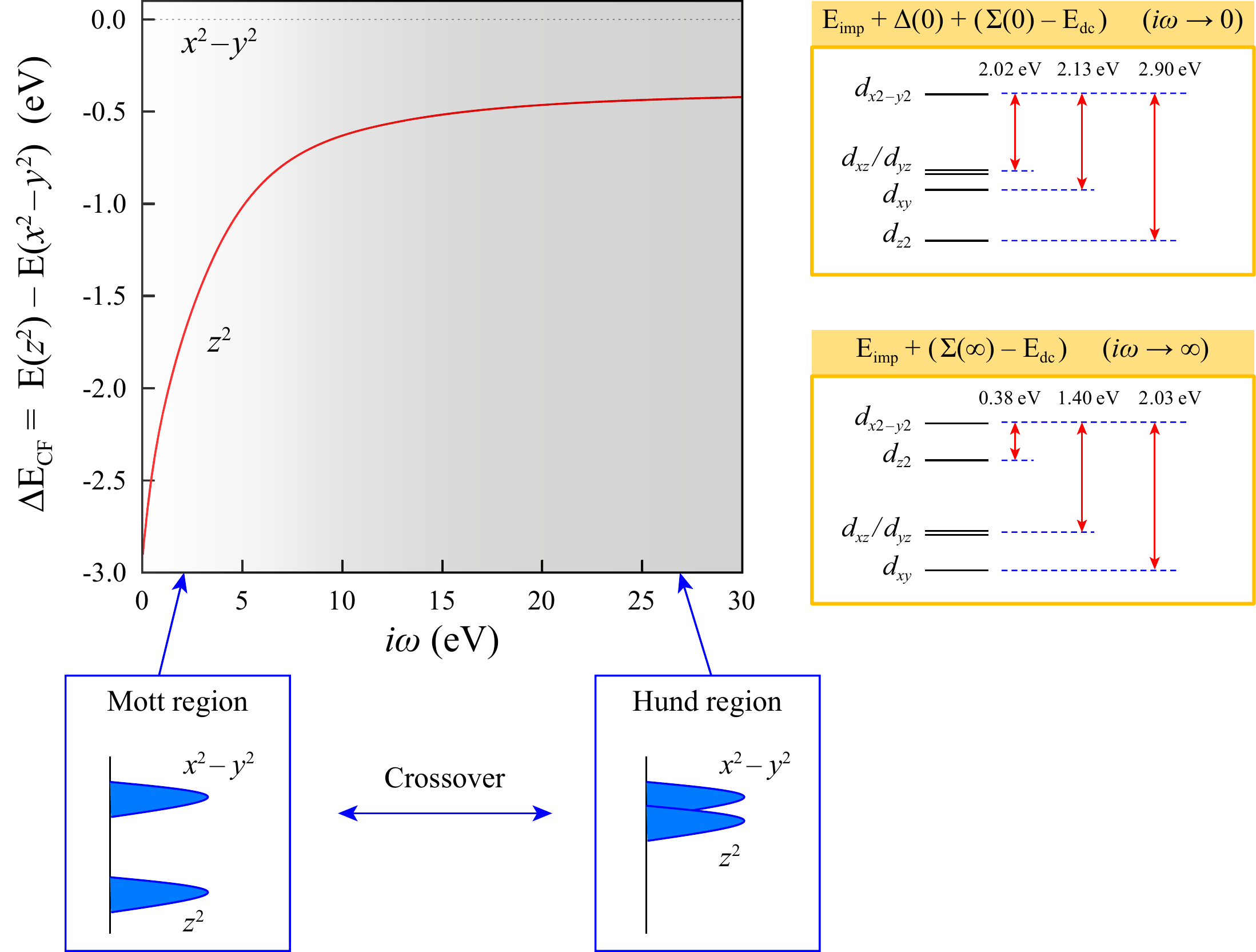}
\caption{Dynamical crystal field splitting
in LaNiO$_{2}$ computed by the DFT+DMFT method.
(Left) Dynamical crystal field splitting of Ni $e_g$ orbitals
with reference energy of the correlated $d_{x^2-y^2}$ orbital.
There is the competition between the crystal field splitting and Hund's coupling
within the Ni $e_g$ orbitals.
A crossover from Mott (white area) to Hund region (gray area) happens
when the dynamical crystal field splitting is comparable
to the strength of $J_{H} \sim$ 1 eV,
where Hund's coupling becomes dominant.
(Right-top) Crystal field splitting of Ni $3d$ orbitals at the zero frequency.
(Right-bottom) Crystal field splitting of Ni $3d$ orbitals at the infinite frequency.
The crystal field levels are arranged in an energy order.
}
\label{fig:Eimp}
\end{figure*}

\begin{figure*}
\centering
\includegraphics[width=0.7\textwidth]{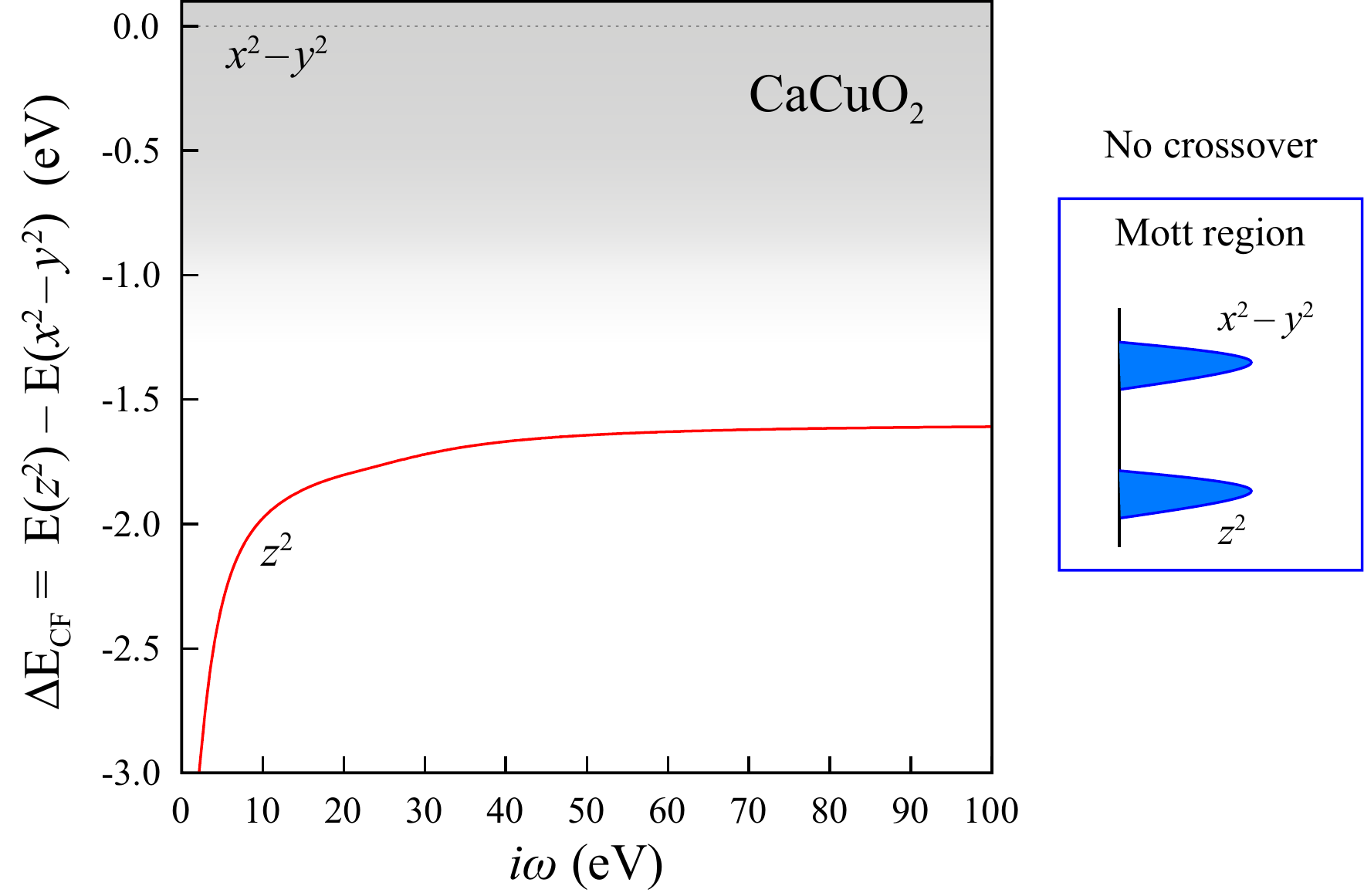}
\caption{Dynamical crystal field splitting of $e_g$ orbitals
in CaCuO$_{2}$ with reference energy of a $d_{x^2-y^2}$ orbital.
It is computed by the DFT+DMFT method
(the Coulomb parameters $U$ = 10 eV and $J_{H}$ = 1 eV are chosen).
There is the competition between the crystal field splitting and Hund's coupling
within the Cu $e_g$ orbitals.
The gray region presents that Hund's coupling becomes more dominant
than the dynamical crystal field splitting.
It clearly shows that there is no crossover from Mott to Hund region
in CaCuO$_{2}$.
}
\label{fig:CaCuO2-Eimp}
\end{figure*}

Here we use the two-band model of Section~\ref{sec:two-band}
to provide a brief qualitative understanding
of the variation of the Hall coefficient $R_{H}$ with doping and temperature.
The correlated $d_{x^2-y^2}$ band displays a coherence-incoherence crossover,
whereas the uncorrelated hybridized band does not.
This is reflected in the strong differentiation of the scattering rates
(see Fig.~\ref{fig:orbital}(b)).
These two bands give the hole and electron carrier, respectively
(see Fig.~\ref{fig:ndnio2-dmft}(a)).
At high temperatures, the correlated $d_{x^2-y^2}$ band is incoherent,
so that the electron carrier from the hybridized band is dominant
and $R_{H}$ shows the negative sign accordingly.
Upon cooling, the correlated $d_{x^2-y^2}$ band gets coherent
and the hole contribution becomes larger.
For $x \leq 0.15$,
the hole contribution is not large enough to compensate the electron contribution
and $R_{H}$ is negative at all temperatures.
For $x \geq 0.18$,
sufficient hole doping makes the hole contribution dominant at low temperatures,
but with rising temperature the hole contribution becomes smaller
(as the correlated band becomes incoherent)
and $R_{H}$ changes sign.

\begin{figure*}
\centering
\includegraphics[width=0.6\textwidth]{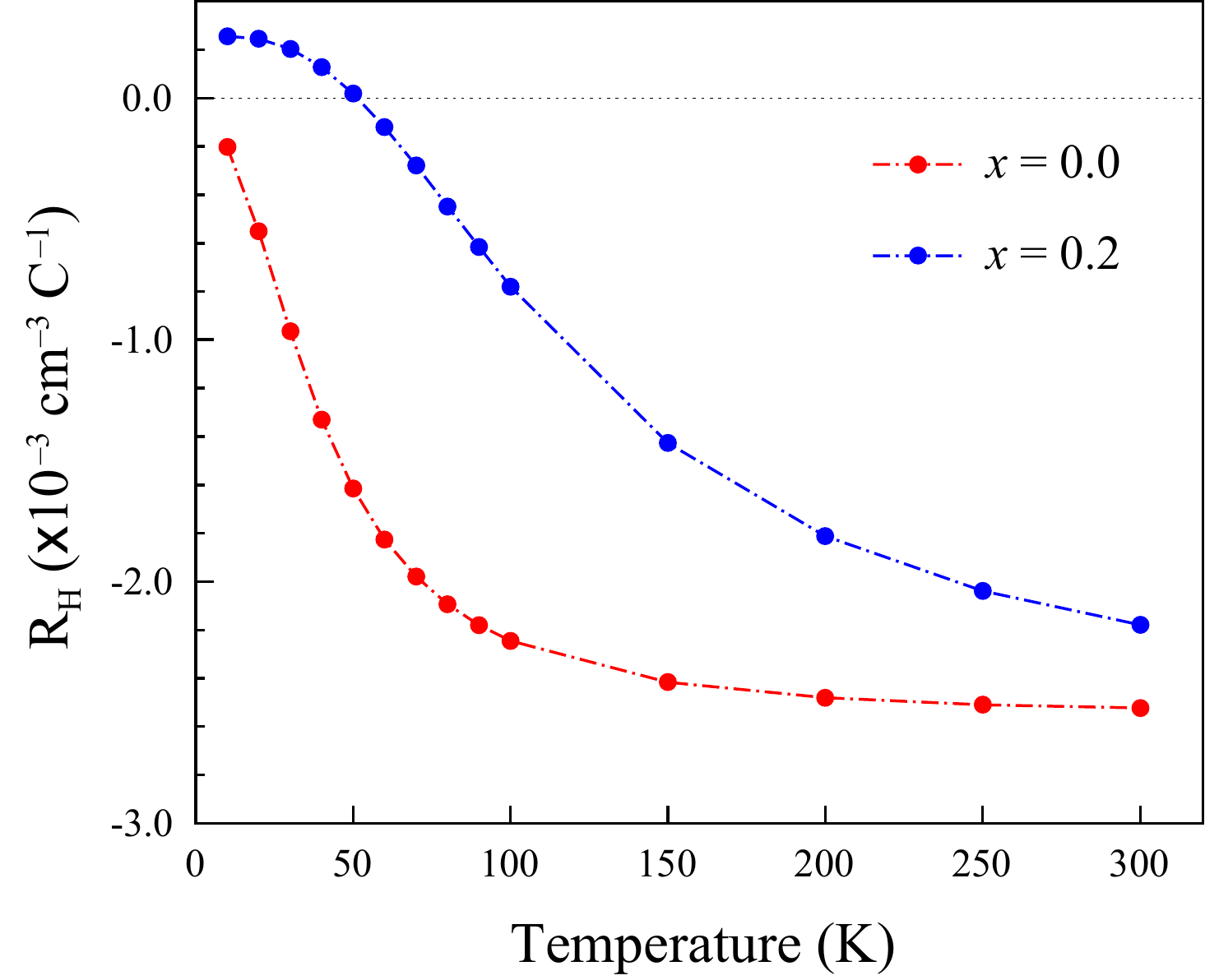}
\caption{Normal state Hall coefficient $R_{H}$
of La$_{1-x}$Sr$_{x}$NiO$_{2}$
computed by Boltzmann transport theory calculations
with the DMFT scattering rates $1/\tau_{qp} = aT^2 + b$
provided in Table~\ref{tab:fitting}.
The order of magnitude is comparable to
the experiments~\cite{Hwang2019,Li2020-arXiv}.
}
\label{fig:Hall}
\end{figure*}

This perspective is confirmed
by Boltzmann transport theory calculations for the two-band model discussed above
using the DMFT scattering rates of Fig.~\ref{fig:orbital}(b)
for the correlated and uncorrelated bands.
Figure~\ref{fig:Hall} shows the computed normal state
Hall coefficients of La$_{1-x}$Sr$_{x}$NiO$_{2}$ for $x$ = 0 and 0.2.
At $x = 0$, $R_{H}$ shows a negative sign at all temperature.
At $x = 0.2$, $R_{H}$ shows a sign change at $\sim$50 K.
Notice that transport calculations
using the same constant relaxation time for the two bands
give a positive sign in $R_{H}$ at all temperature for any doping concentration $x$.

All this is in qualitative agreement with experiments~\cite{Hwang2019,Zeng2020-arXiv,Li2020-arXiv},
on the normal state $R_{H}$ of  Nd$_{1-x}$Sr$_{x}$NiO$_{2}$.
It shows a negative sign at all temperatures for $x \leq 0.15$,
while it undergoes a sign change for $x \geq 0.18$.
For $x$ = 0.2, the sign change happens at $\sim$50 K.
These aspects of the data are in qualitative agreement with our results.
More quantitative description of the transport including
the weak localization effects at low doping and low temperatures
requires the inclusion of disorder effects which is outside the
scope of this paper.


\begin{thebibliography}{99}

\bibitem{Hwang2019}
    Danfeng Li, Kyuho Lee, Bai Yang Wang, Motoki Osada, Samuel Crossley, Hye Ryoung Lee, Yi Cui, Yasuyuki Hikita, and Harold Y. Hwang,
    ``Superconductivity in an infinite-layer nickelate,''
    \href{https://doi.org/10.1038/s41586-019-1496-5}
    {Nature {\bf 572}, 624 (2019)}.


\bibitem{Lee2020-apl}
    Kyuho Lee, Berit H. Goodge, Danfeng Li, Motoki Osada, Bai Yang Wang, Yi Cui, Lena F. Kourkoutis, and Harold Y. Hwang,
    ``Aspects of the synthesis of thin film superconducting infinite-layer nickelates,''
    \href{https://doi.org/10.1063/5.0005103}
    {APL Materials {\bf 8}, 041107 (2020)}.


\bibitem{Osada2020-nano}
    Motoki Osada, Bai Yang Wang, Berit H. Goodge, Kyuho Lee, Hyeok Yoon, Keita Sakuma, Danfeng Li, Masashi Miura, Lena F. Kourkoutis, and Harold Y. Hwang,
    ``A superconducting praseodymium nickelate with infinite layer structure,''
    \href{https://doi.org/10.1021/acs.nanolett.0c01392}
    {Nano Lett. {\bf 20}, 5735 (2020)}.


\bibitem{Zeng2020-arXiv}
    Shengwei Zeng, Chi Sin Tang, Xinmao Yin, Cgangjian Li, Zhen Huang, Junxiong Hu, Wei Liu, Ganesh Ji Omar, Hariom Jani, Zhi Shiuh Lim, Kun Han, Dongyang Wan, Ping Yang, Andrew T. S. Wee, Ariando Ariando,
    ``Phase diagram and superconducting dome of infinte-layer Nd$_{1-x}$Sr$_{x}$NiO$_{2}$ thin films,''
    \href{https://doi.org/10.1103/PhysRevLett.125.147003}
    {Phys. Rev. Lett. {\bf 125}, 147003 (2020)}.


\bibitem{Li2020-arXiv}
    Danfeng Li, Bai Yang Wang, Kyuho Lee, Shannon P. Harvey, Motoki Osada, Berit H. Goodge, Lena F. Kourkoutis, Harold Y. Hwang,
    ``Superconducting Dome in Nd$_{1-x}$Sr$_{x}$NiO$_{2}$ Infinite Layer Films,''
    \href{https://doi.org/10.1103/PhysRevLett.125.027001}
    {Phys. Rev. Lett. {\bf 125}, 027001 (2020)}.


\bibitem{Gu2020-stm}
    Qiangqiang Gu, Yueying Li, Siyuan Wan, Huazhou Li, Wei Guo, Huan Yang, Qing Li, Xiyu Zhu, Xiaoqing Pan, Yuefeng Nie, and Hai-Hu Wen,
    ``Single particle tunneling spectrum of superconducting
    Nd$_{1-x}$Sr$_{x}$NiO$_{2}$ thin films,''
    \href{https://doi.org/10.1038/s41467-020-19908-1}
    {Nat. Commun. {\bf 11}, 6027 (2020)}.



\bibitem{Li2020}
    Qing Li, Chengping He, Jin Si, Xiyu Zhu, and Hai-Hu Wen,
    ``Absence of superconductivity in bulk Nd$_{1-x}$Sr$_{x}$NiO$_{2}$,''
    \href{https://doi.org/10.1038/s43246-020-0018-1}
    {Communications Materials {\bf 1}, 16 (2020)}.


\bibitem{Zhou2020}
    Xiao-Rong Zhou, Ze-Xin Feng, Pei-Xin Qin, Han Yan, Shuai Hu, Hui-Xin Guo, Xiao-Ning Wang, Hao-Jiang Wu, Xin Zhang, Hong-Yu Chen, Xue-Peng Qiu, and Zhi-Qi Liu,
    ``Absence of superconductivity in Nd$_{0.8}$Sr$_{0.2}$NiO$_{x}$ thin films without chemical reduction,''
    \href{https://doi.org/10.1007/s12598-020-01389-2}
    {Rare Metals {\bf 39}, 368 (2020)}.



\bibitem{Crespin1983}
    Michel Crespin, Pierre Levitz, and Lucien Gatineau,
    ``Reduced forms of LaNiO$_{3}$ perovskite. Part 1.---Evidence for new phases: La$_{2}$Ni$_{2}$O$_{5}$ and LaNiO$_{2}$,''
    \href{https://doi.org/10.1039/F29837901181}
    {J. Chem. Soc., Faraday Trans. 2 {\bf 79}, 1181 (1983)}.

\bibitem{Pierre1983}
    Pierre Levitz, Michel Crespin, and Lucien Gatineau,
    ``Reduced forms of LaNiO$_{3}$ perovskite. Part 2.---X-ray structure of LaNiO$_{2}$ and extended X-ray absorption fine structure study: local environment of monovalent nickel,''
    \href{https://doi.org/10.1039/F29837901195}
    {J. Chem. Soc., Faraday Trans. 2 {\bf 79}, 1195 (1983)}.

\bibitem{Hayward1999-jacs}
    M. A. Hayward, M. A. Green, M. J. Rosseinsky, and J. Sloan,
    ``Sodium Hydride as a Powerful Reducing Agent for Topotactic Oxide Deintercalation: Synthesis and Characterization of the Nickel(I) Oxide LaNiO$_{2}$,''
    \href{https://doi.org/10.1021/ja991573i}
    {J. Am. Chem. Soc. {\bf 121}, 8843 (1999)}.

\bibitem{Ikeda2016}
    Ai Ikeda, Yoshihara Krockenberger, Hiroshi Irie, Michio Naito, and Hideki Yamamoto,
    ``Direct observation of infinite NiO$_{2}$ planes in LaNiO$_{2}$ films,''
    \href{http://doi.org/10.7567/APEX.9.061101}
    {Applied Physics Express {\bf 9}, 061101 (2016)}.


\bibitem{Ikeda2013}
    Ai Ikeda, Takaaki Manabe, Michio Naito,
    ``Improved conductivity of infite-layer LaNiO$_{2}$ thin films by metal organic decomposition,''
    \href{https://doi.org/10.1016/j.physc.2013.09.007}
    {Physica C {\bf 495}, 134 (2013)}.


\bibitem{Lee2020-nm}
    M. Hepting, D. Li, C. J. Jia, H. Lu, E. Paris, Y. Tseng, X. Feng, M. Osada, E. Been, Y. Hikita, Y.-D. Chuang, Z. Hussain, K. J. Zhou, A. Nag, M. Garcia-Fernandez, M. Rossi, H. Y. Huang, D. J. Huang, Z. X. Shen, T. Schmitt, H. Y. Hwang, B. Moritz, J. Zaanen, T. P. Devereaux, and W. S. Lee,
    ``Electronic structure of the parent compound of superconducting infinite-layer nickelates,''
    \href{https://doi.org/10.1038/s41563-019-0585-z}
    {Nature Materials {\bf 19}, 381 (2020)}.

\bibitem{Mei2019}
    Ying Fu, Le Wang, Hu Cheng, Shenghai Pei, Xuefeng Zhou, Jian Chen, Shaoheng Wang, Ran Zhao, Wenrui Jiang, Cai Liu, Mingyuan Huang, XinWei Wang, Yusheng Zhao, Dapeng Yu, Fei Ye, Shanmin Wang, Jia-Wei Mei,
    ``Core-level x-ray photoemission and Raman spectroscopy studies on electronic structures in Mott-Hubbard type nickelate oxide NdNiO$_{2}$,''
    \href{https://arxiv.org/abs/1911.03177}
    {arXiv:1911.03177 (2019)}.

\bibitem{Goodge2020-arXiv}
    Berit H. Goodge, Danfeng Li, Motoki Osada, Bai Yang Wang, Kyuho Lee, George A. Sawatzky, Harold Y. Hwang, and Lena F. Korkoutis,
    ``Doping evolution of the Mott-Hubbard landscape in infinite-layer nickelates,''
    \href{https://arxiv.org/abs/2005.02847}
    {arXiv:2005.02847 (2020)}.


\bibitem{Nomura2019}
    Yusuke Nomura, Motoaki Hirayama, Terumasa Tadano, Yoshihide Yoshimoto, Kazuma Nakamura, and Ryotaro Arita,
    ``Formation of a two-dimensional single-component correlated electron system and band engineering in the nickelate superconductor NdNiO$_{2}$,''
    \href{https://doi.org/10.1103/PhysRevB.100.205138}
    {Phys. Rev. B {\bf 100}, 205138 (2019)}.

\bibitem{Pickett2004}
    K.-W. Lee and W. E. Pickett,
    ``Infinite-layer LaNiO$_{2}$: Ni$^{1+}$ is not Cu$^{2+}$,''
    \href{https://doi.org/10.1103/PhysRevB.70.165109}
    {Phys. Rev. B {\bf 70}, 165109 (2004)}.

\bibitem{Norman2020-prx}
    A. S. Botana and M. R. Norman,
    ``Similarities and Differences between LaNiO$_{2}$ and CaCuO$_{2}$ and Implications for Superconductivity,''
    \href{https://doi.org/10.1103/PhysRevX.10.011024}
    {Phys. Rev. X {\bf 10}, 011024 (2020)}.

\bibitem{Hirayama2020}
    Motoaki Hirayama, Terumasa Tadano, Yusuke Nomura, and Ryotaro Arita,
    ``Materials design of dynamically stable $d^9$ layered nickelates,''
    \href{https://doi.org/10.1103/PhysRevB.101.075107}
    {Phys. Rev. B {\bf 101}, 075107 (2020)}.

\bibitem{Zhang2020-prr}
    Hu Zhang, Lipeng Jin, Shanmin Wang, Bin Xi, Xingqiang Shi, Fei Ye, and Jia-Wei Mei,
    ``Effective Hamiltonian for nickelate oxides Nd$_{1-x}$Sr$_{x}$NiO$_{2}$,''
    \href{https://doi.org/10.1103/PhysRevResearch.2.013214}
    {Phys. Rev. Research {\bf 2}, 013214 (2020)}.

\bibitem{Lechermann2020}
    Frank Lechermann,
    ``Late transition metal oxides with infinite-layer structure: Nickelates versus cuprates,''
    \href{https://doi.org/10.1103/PhysRevB.101.081110}
    {Phys. Rev. B {\bf 101}, 081110(R) (2020)}.

\bibitem{Karp2020-prx}
    Jonathan Karp, Antia S. Botana, Michael R. Norman, Hyowon Park, Manuel Zingl, and Andrew Millis,
    ``Many-Body Electronic Structure of NdNiO$_{2}$ and CaCuO$_{2}$,''
    \href{https://doi.org/10.1103/PhysRevX.10.021061}
    {Phys. Rev. X {\bf 10}, 021061 (2020)}.

\bibitem{Kitatani2020-arXiv}
    Motoharu Kitatani, Liang Si, Oleg Janson, Ryotaro Arita, Zhicheng Zhong, Karsten Held,
    ``Nickelate superconductors - a renaissance of the one-band Hubbard model,''
    \href{https://doi.org/10.1038/s41535-020-00260-y}
    {npj Quantum Mater. {\bf 5}, 59 (2020)}.

\bibitem{Thomale2020}
    Xianxin Wu, Domenico Di Sante, Tilman Schwemmer, Werner Hanke, Harold Y. Hwang, Srinivas Raghu, and Ronny Thomale,
    ``Robust $d_{x^2-y^2}$-wave superconductivity of infinite-layer nickelates,''
    \href{https://doi.org/10.1103/PhysRevB.101.060504}
    {Phys. Rev. B {\bf 101}, 060504(R) (2020)}.

\bibitem{Sawatzky2020-prl}
    Mi Jiang, Mona Berciu, and George A. Sawatzky,
    ``Critical Nature of the Ni Spin State in Doped NdNiO$_{2}$,''
    \href{https://doi.org/10.1103/PhysRevLett.124.207004}
    {Phys. Rev. Lett. {\bf 124}, 207004 (2020)}.

\bibitem{YilinWang2020-arXiv}
    Y. Wang, C.-J. Kang, H. Miao, G. Kotliar,
    ``Hund's metal physics: from SrNiO$_{2}$ to NdNiO$_{2}$,''
    \href{https://doi.org/10.1103/PhysRevB.102.161118}
    {Phys. Rev. B {\bf 102}, 161118(R) (2020)}.

\bibitem{Choi2020-arXiv}
    Byungkyun Kang, Corey Melnick, Patrick Semon, Gabriel Kotliar, Sangkook Choi,
    ``Infinite-layer nickelates as Ni-eg Hund's metals,''
    \href{https://arxiv.org/abs/2007.14610}
    {arXiv:2007.14610 (2020)}.

\bibitem{Werner2020-arXiv}
    Francesco Petocchi, Viktor Christiansson, Fredrik Nilsson, Ferdi Aryasetiawan, and Philipp Werner,
    ``Normal state of Nd$_{1-x}$Sr$_{x}$NiO$_{2}$ from self-consistent GW+EDMFT,''
    \href{https://arxiv.org/abs/2006.00394}
    {arXiv:2006.00394 (2020)}.

\bibitem{Hu2019-prr}
    Lun-Hui Hu and Congjun Wu,
    ``Two-band model for magnetism and superconductivity in nickelates,''
    \href{https://doi.org/10.1103/PhysRevResearch.1.032046}
    {Phys. Rev. Research {\bf 1}, 032046(R) (2019)}.

\bibitem{Werner2020-prb}
    Philipp Werner and Shintaro Hoshino,
    ``Nickelate superconductors: Multiorbital nature and spin freezing,''
    \href{https://doi.org/10.1103/PhysRevB.101.041104}
    {Phys. Rev. B {\bf 101}, 041104(R) (2020)}.

\bibitem{Vishwanath2020}
    Ya-Hui Zhang and Ashvin Vishwanath,
    ``Type-II $t-J$ model in superconducting nickelate Nd$_{1-x}$Sr$_{x}$NiO$_{2}$,''
    \href{https://doi.org/10.1103/PhysRevResearch.2.023112}
    {Phys. Rev. Research {\bf 2}, 023112 (2020)}.

\bibitem{Chang2019-arXiv}
    Jun Chang, Jize Zhao, Yang Ding,
    ``Hund-Heisenberg model in superconducting infinite-layer nickelates,''
    \href{https://doi.org/10.1140/epjb/e2020-10343-7}
    {Phys. J. B {\bf 93}, 220 (2020)}.

\bibitem{Been2020-arXiv}
    Emily Been, Wei-Sheng Lee, Harold Y. Hwang, Yi Cui, Jan Zaanen, Thomas Devereaux, Brian Moritz, Chunjing Jia,
    ``Theory of Rare-earth Infinite Layer Nickelates,''
    \href{https://arxiv.org/abs/2002.12300}
    {arXiv:2002.12300 (2020)}.

\bibitem{Lechermann2020-arXiv}
    Frank Lechermann,
    ``Multiorbital Processes Rule the Nd$_{1-x}$Sr$_{x}$NiO$_{2}$ Normal State,''
    \href{https://doi.org/10.1103/PhysRevX.10.041002}
    {Phys. Rev. X {\bf 10}, 041002 (2020)}.

\bibitem{Gu2020}
    Yuhao Gu, Sichen Zhu, Xiaoxuan Wang, Jiangping Hu, and Hanghui Chen,
    ``A substantial hybridization between correlated Ni-$d$ orbital and itinerant electrons in infinite-layer nickelates,''
    \href{https://doi.org/10.1038/s42005-020-0347-x}
    {Communications Physics {\bf 3}, 84 (2020)}.

\bibitem{Zhang2020-Kondo}
    Guang-Ming Zhang, Yi-feng Yang, and Fu-Chun Zhang,
    ``Self-doped Mott insulator for parent compounds of nickelate superconductors,''
    \href{https://doi.org/10.1103/PhysRevB.101.020501}
    {Phys. Rev. B {\bf 101}, 020501(R) (2020)}.

\bibitem{Wang2020-arXiv}
    Zhan Wang, Guang-Ming Zhang, Yi-feng Yang, and Fu-Chun Zhang,
    ``Distinct pairing symmetries of superconductivity in infinite-layer nickelates,''
    \href{https://doi.org/10.1103/PhysRevB.102.220501}
    {Phys. Rev. B {\bf 102}, 220501(R) (2020)}.

\bibitem{Jiang2019}
    Peiheng Jiang, Liang Si, Zhaoliang Liao, and Zhicheng Zhong,
    ``Electronic structure of rare-earth infinite-layer $R$NiO$_{2}$ ($R$ = La, Nd),''
    \href{https://doi.org/10.1103/PhysRevB.100.201106}
    {Phys. Rev. B {\bf 100}, 201106(R) (2019)}.

\bibitem{Held2020}
    Liang Si, Wen Xiao, Josef Kaufmann, Jan M. Tomczak, Yi Lu, Zhicheng Zhong, and Karsten Held,
    ``Topotactic Hydrogen in Nickelate Superconductors and Akin Infinite-Layer Oxides $AB$O$_{2}$,''
    \href{https://doi.org/10.1103/PhysRevLett.124.166402}
    {Phys. Rev. Lett. {\bf 124}, 166402 (2020)}.

\bibitem{Ryee2020}
    Siheon Ryee, Hongkee Yoon, Taek Jung Kim, Min Yong Jeong, and Myung Joon Han,
    ``Induced magnetic two-dimensionality by hole doping in the superconducting infinite-layer nickelate Nd$_{1-x}$Sr$_{x}$NiO$_{2}$,''
    \href{https://doi.org/10.1103/PhysRevB.101.064513}
    {Phys. Rev. B {\bf 101}, 064513 (2020)}.

\bibitem{Savrasov2020-prb}
    I. Leonov, S. L. Skornyakov, and S. Y. Savrasov,
    ``Lifshitz transition and frustration of magnetic moments in infinite-layer NdNiO$_{2}$ upon hole doping,''
    \href{https://doi.org/10.1103/PhysRevB.101.241108}
    {Phys. Rev. B {\bf 101}, 241108(R) (2020)}.

\bibitem{Sakakibara2019-arXiv}
    Hirofumi Sakakibara, Hidetomo Usui, Katsuhiro Suzuki, Takao Kotani, Hideo Aoki, Kazuhiko Kuroki,
    ``Model Construction and a Possibility of Cupratelike Pairing in a New $d^9$ Nickelate Superconductor (Nd,Sr)NiO$_{2}$,''
    \href{https://doi.org/10.1103/PhysRevLett.125.077003}
    {Phys. Rev. Lett. {\bf 125}, 077003 (2020)}.

\bibitem{Adhikary2020-arXiv}
    Priyo Adhikary, Subhadeep Bandyopadhyay, Tanmoy Das, Indra Dasgupta, Tanusri Saha-Dasgupta,
    ``Orbital-selective superconductivity in a two-band model of infinite-layer nickelates,''
    \href{https://doi.org/10.1103/PhysRevB.102.100501}
    {Phys. Rev. B {\bf 102}, 100501(R) (2020)}.

\bibitem{Pickett2020-arXiv}
    Mi-Young Choi, W. E. Pickett, K.-W. Lee,
    ``Quantum-Fluctuation-Frustrated Flat Band Instabilities in NdNiO$_{2}$,''
    \href{https://arxiv.org/abs/2005.03234v1}
    {arXiv:2005.03234 (2020)}.

\bibitem{Cano2020-GW}
    Valerio Olevano, Fabio Bernardini, Xavier Blase, Andr\'{e}s Cano,
    ``Ab initio many-body GW correlations in the electronic structure of LaNiO$_{2}$,''
    \href{https://arxiv.org/abs/2001.09194v1}
    {arXiv:2001.09194 (2020)}.

\bibitem{Geisler2020-arXiv}
    Benjamin Geisler and Rossitza Pentcheva,
    ``Fundamental difference in the electronic reconstruction of infinite-layer versus perovskite neodymium nickelate films on SrTiO$_{3}$(001),''
    \href{https://doi.org/10.1103/PhysRevB.102.020502}
    {Phys. Rev. B {\bf 102}, 020502(R) (2010)}.

\bibitem{Bernardini2020}
    Fabio Bernardini and Andres Cano,
    ``Stability and electronic properties of LaNiO$_{2}$/SrTiO$_{3}$ heterostructures,''
    \href{https://doi.org/10.1088/2515-7639/ab9d0f}
    {J. Phys. Mater. {\bf 3}, 03LT01 (2020)}.

\bibitem{Gao2020-arXiv}
    Jiacheng Gao, Zhijun Wang, Chen Fang, Hongming Weng,
    ``Electronic structures and topological properties in nickelates $Ln_{n+1}$Ni$_{n}$O$_{2n+2}$,''
    \href{https://doi.org/10.1093/nsr/nwaa218}
    {National Science Review, nwaa218 (2020)}.



\bibitem{Georges1996}
    Antoine Georges, Gabriel Kotliar, Werner Krauth, and Marcelo J. Rozenberg,
    ``Dynamical mean-field theory of strongly correlated fermion systems and the limit of infinite dimensions,''
    \href{https://doi.org/10.1103/RevModPhys.68.13}
    {Rev. Mod. Phys. {\bf 68}, 13 (1996)}.


\bibitem{Kotliar2006}
    G. Kotliar, S. Y. Savrasov, K. Haule, V. S. Oudovenko, O. Parcollet, and C. A. Marianetti,
    ``Electronic structure calculations with dynamical mean-field theory,''
    \href{https://doi.org/10.1103/RevModPhys.78.865}
    {Rev. Mod. Phys. {\bf 78}, 865 (2006)}.


\bibitem{Held2007}
    K. Held,
    ``Electronic structure calculations using dynamical mean field theory,''
    \href{https://doi.org/10.1080/00018730701619647}
    {Advances in Physics {\bf 56}, 829 (2007)}.



\bibitem{Haule2010-prb}
    Kristjan Haule, Chuck-Hou Yee, and Kyoo Kim,
    ``Dynamical mean-field theory within the full-potential methods: Electronic structure of CeIrIn$_{5}$, CeCoIn$_{5}$, and CeRhIn$_{5}$,''
    \href{https://doi.org/10.1103/PhysRevB.81.195107}
    {Phys. Rev. B {\bf 81}, 195107 (2010)}.

\bibitem{Wien2k}
    Peter Blaha, Karlheinz Schwarz, Fabien Tran, Robert Laskowski, Georg K. H. Madsen, and Laurence D. Marks,
    ``WIEN2k: An APW+lo program for calculating the properties of solids,''
    \href{https://doi.org/10.1063/1.5143061}
    {J. Chem. Phys. {\bf 152}, 074101 (2020)}.

\bibitem{Haule2015}
    Kristjan Haule,
    ``Exact Double Counting in Combining the Dynamical Mean Field Theory and the Density Functional Theory,''
    \href{https://doi.org/10.1103/PhysRevLett.115.196403}
    {Phys. Rev. Lett. {\bf 115}, 196403 (2015)}.


\bibitem{Suppl}
    See Supplemental Material, which includes Refs.~\cite{Hwang2019,Zeng2020-arXiv,Li2020-arXiv,
    Wien2k,Haule2010-prb,Hayward1999-jacs,Werner2006,Haule2007-ctqmc,Perdew1996,
    Haule2015,Jarrell1996-maxent,Deng2014,Pickett2020-prb,wannier90-1,wannier90-2,
    Mravlje2011,Mravlje2016,Haule2009-njp,Hardy2013,Miao2016,Liu2015,
    Sakakibara2019-arXiv,Nomura2019,Lee2020-nm,Adhikary2020-arXiv,
    Wan2020-arXiv,Qazilbash2009},
    for the computational details, DFT electronic structure, decomposition of the Drude peak, impact of broadening factor on optic calculations, comparison with NdNiO$_{2}$,
    optical conductivity for different Coulomb $U$ parameters, comparison with CaCuO$_{2}$, optics for a prototypical Mott system of V$_{2}$O$_{3}$, 3-dimensional Fermi surfaces of La$_{1-x}$Sr$_{x}$NiO$_{2}$ computed with DFT+DMFT, the DMFT valence histogram as a function of Hund's coupling $J_{H}$, coherence-incoherence crossover, the two-band model based on the Wannier interpretation, and the dynamical crystal field splitting.


\bibitem{Basov2011-rmp}
    D. N. Basov, Richard D. Averitt, Dirk van der Marel, Martin Dressel, and Kristjan Haule,
    ``Electrodynamics of correlated electron materials,''
    \href{https://doi.org/10.1103/RevModPhys.83.471}
    {Rev. Mod. Phys. {\bf 83}, 471 (2011)}.

\bibitem{Qazilbash2009}
    M. M. Qazilbash, J. J. Hamlin, R. E. Baumbach, Lijun Zhang, D. J. Singh, M. B. Maple, and D. N. Basov,
    ``Electronic correlations in the iron pnictides,''
    \href{https://doi.org/10.1038/nphys1343}
    {Nature Physics {\bf 5}, 647 (2009)}.

\bibitem{Deng2014}
    Xiaoyu Deng, Aaron Sternbach, Kristjan Haule, D. N. Basov, and Gabriel Kotliar,
    ``Shining Light on Transition-Metal Oxides: Unveiling the Hidden Fermi Liquid,''
    \href{https://doi.org/10.1103/PhysRevLett.113.246404}
    {Phys. Rev. Lett. {\bf 113}, 246404 (2014)}.

\bibitem{Werner2006}
    Philipp Werner, Armin Comanac, Luca de' Medici, Matthias Troyer, and Andrew J. Millis,
    ``Continuous-Time Solver for Quantum Impurity Models,''
    \href{https://doi.org/10.1103/PhysRevLett.97.076405}
    {Phys. Rev. Lett. {\bf 97}, 076405 (2006)}.

\bibitem{Haule2007-ctqmc}
    Kristjan Haule,
    ``Quantum Monte Carlo impurity solver for cluster dynamical mean-field theory and electronic structure calculations with adjustable cluster base,''
    \href{https://doi.org/10.1103/PhysRevB.75.155113}
    {Phys. Rev. B {\bf 75}, 155113 (2007)}.

\bibitem{Perdew1996}
    John P. Perdew, Kieron Burke, and Matthias Ernzerhof,
    ``Generalized Gradient Approximation Made Simple,''
    \href{https://doi.org/10.1103/PhysRevLett.77.3865}
    {Phys. Rev. Lett. {\bf 77}, 3865 (1996)}.

\bibitem{Jarrell1996-maxent}
    Mark Jarrell, J. E. Gubernatis,
    ``Bayesian inference and the analytic continuation of imaginary-time quantum Monte Carlo data,''
    \href{https://doi.org/10.1016/0370-1573(95)00074-7}
    {Physics Reports {\bf 269}, 133 (1996)}.

\bibitem{wannier90-1}
    Arash A. Mostofi, Jonathan R. Yates, Young-Su Lee, Ivo Souza, David Vanderbilt, Nicola Marzari,
    ``wannier90: A tool for obtaining maximally-localised Wannier functions,''
    \href{https://doi.org/10.1016/j.cpc.2007.11.016}
    {Computer Physics Communications {\bf 178}, 685 (2008)}.

\bibitem{wannier90-2}
    Giovanni Pizzi {\it et al.},
    ``Wannier90 as a community code: new features and applications,''
    \href{https://doi.org/10.1088/1361-648X/ab51ff}
    {J. Phys.: Condens. Matter {\bf 32}, 165902 (2020)}.




\bibitem{Pickett2020-prb}
    Mi-Young Choi, Kwan-Woo Lee, and Warren E. Pickett,
    ``Role of $4f$ states in infinite-layer NdNiO$_{2}$''
    \href{https://doi.org/10.1103/PhysRevB.101.020503}
    {Phys. Rev. B {\bf 101}, 020503(R) (2020)}.

\bibitem{Wan2020-arXiv}
    Xiangang Wan, Vsevolod Ivanov, Giacomo Resta, Ivan Leonov, Sergey Y. Savrasov,
    ``Calculated Exchange Interactions and Competing $S$ = 0 and $S$ = 1 States in Doped NdNiO$_{2}$''
    \href{https://arxiv.org/abs/2008.07465}
    {arXiv:2008.07465 (2020)}.


\bibitem{Haule2009-njp}
    K. Haule and G. Kotliar,
    ``Coherence-incoherence crossover in the normal state of iron oxypnictides and importance of Hund's rule coupling,''
    \href{https://doi.org/10.1088/1367-2630/11/2/025021}
    {New J. Phys. {\bf 11}, 025021 (2009)}.


\bibitem{Mravlje2011}
    Jernej Mravlje, Markus Aichhorn, Takashi Miyake, Kristjan Haule, Gabriel Kotliar, and Antoine Georges,
    ``Coherence-Incoherence Crossover and the Mass-Renormalization Puzzles in Sr$_{2}$RuO$_{4}$,''
    \href{https://doi.org/10.1103/PhysRevLett.106.096401}
    {Phys. Rev. Lett. {\bf 106}, 096401 (2011)}.


\bibitem{Mravlje2016}
    Jernej Mravlje and Antoine Georges,
    ``Thermopower and Entropy: Lessons from Sr$_{2}$RuO$_{4}$,''
    \href{https://doi.org/10.1103/PhysRevLett.117.036401}
    {Phys. Rev. Lett. {\bf 117}, 036401 (2016)}.


\bibitem{Hardy2013}
    F. Hardy, A. E. B\"{o}hmer, D. Aoki, P. Burger, T. Wolf, P. Schweiss, R. Heid, P. Adelmann, Y. X. Yao, G. Kotliar, J. Schmalian, and C. Meingast,
    ``Evidence of Strong Correlations and Coherence-Incoherence Crossover in the Iron Pnictide Superconductor KFe$_{2}$As$_{2}$,''
    \href{https://doi.org/10.1103/PhysRevLett.111.027002}
    {Phys. Rev. Lett. {\bf 111}, 027002 (2013)}.


\bibitem{Miao2016}
    H. Miao, Z. P. Yin, S. F. Wu, J. M. Li, J. Ma, B.-Q. Lv, X. P. Wang, T. Qian, P. Richard, L.-Y. Xing, X.-C. Wang, C. Q. Jin, K. Haule, G. Kotliar, and H. Ding,
    ``Orbital-differentiated coherence-incoherence crossover identified by photoemisson spectroscopy in LiFeAs,''
    \href{https://doi.org/10.1103/PhysRevB.94.201109}
    {Phys. Rev. B {\bf 94}, 201109(R) (2016)}.


\bibitem{Liu2015}
    Z. K. Liu, M. Yi, Y. Zhang, J. Hu, R. Yu, J.-X. Zhu, R.-H. He, Y. L. Chen, M. Hashimoto, R. G. Moore, S.-K. Mo, Z. Hussain, Q. Si, Z. Q. Mao, D. H. Lu, and Z.-X. Shen,
    ``Experimental observation of incoherent-coherent crossover and orbital-dependent band renormalization in iron chalcogenide superconductors,''
    \href{https://doi.org/10.1103/PhysRevB.92.235138}
    {Phys. Rev. B {\bf 92}, 235138 (2015)}.





\bibitem{Schafgans2012}
    A. A. Schafgans, S. J. Moon, B. C. Pursley, A. D. LaForge, M. M. Qazilbash, A. S. Sefat, D. Mandrus, K. Haule, G. Kotliar, and D. N. Basov,
    ``Electronic Correlations and Unconventional Spectral Weight Transfrer in the High-Temperature Pnictide BaFe$_{2-x}$Co$_{x}$As$_{2}$ Superconductor Using Infrared Spectroscopy,''
    \href{https://doi.org/10.1103/PhysRevLett.108.147002}
    {Phys. Rev. Lett. {\bf 108}, 147002 (2012)}.


\bibitem{U-dependence}
    To provide an additional test for the Hund's character of the LaNiO$_{2}$ system,
    we investigate the Coulomb parameter $U$ dependence of the optical conductivity
    (see Section VI in Supplemental Material~\cite{Suppl}).
    Increasing $U$ just reduces the overall magnitude of the optical conductivity
    (since $U$ suppresses the charge fluctuations)
    but does not affect all the interesting spectral weight transfer effects
    as a function of temperature and doping
    which are the signatures of the Hund's metal.



\bibitem{Z-factor}
    $Z$ is estimated from $1/Z = 1-\partial\text{Im}\Sigma(i\omega) / \partial\omega|_{\omega\rightarrow0^{+}}$
    and it is comparable to the result from analytical continuation of the modified Gaussian method.


\bibitem{Deng2013}
    Xiaoyu Deng, Jernej Mravlje, Rok \v{Z}itko, Michel Ferrero,
    Gabriel Kotliar, and Antoine Georges,
    ``How Bad Metals Turn Good: Spectroscopic Signatures of Resilient Quasiparticles,''
    \href{https://doi.org/10.1103/PhysRevLett.110.086401}
    {Phys. Rev. Lett. {\bf 110}, 086401 (2013)}.

\bibitem{Xu2013}
    Wenhu Xu, Kristjan Haule, and Gabriel Kotliar,
    ``Hidden Fermi Liquid, Scattering Rate Saturation, and Nernst Effect:
    A Dynamical Mean-Field Theory Perspective,''
    \href{https://doi.org/10.1103/PhysRevLett.111.036401}
    {Phys. Rev. Lett. {\bf 111}, 036401 (2013)}.


\bibitem{Deng2016}
    Xiaoyu Deng, Kristjan Haule, and Gabriel Kotliar,
    ``Transport Properties of Metallic Ruthenates: A DFT + DMFT Investigation,''
    \href{https://doi.org/10.1103/PhysRevLett.116.256401}
    {Phys. Rev. Lett. {\bf 116}, 256401 (2016)}.



\bibitem{Imada1998-rmp}
    Masatoshi Imada, Atsushi Fujimori, and Yoshinori Tokura,
    ``Metal-insulator transitions,''
    \href{https://doi.org/10.1103/RevModPhys.70.1039}
    {Rev. Mod. Phys. {\bf 70}, 1039 (1998)}.


\bibitem{occupancy}
    $n_{e}$ for Ni $d_{x^2-y^2}$ changes significantly upon doping,
    however its high-energy occupation number, that is obtained from integration of the DMFT spectral function $A(\omega)$ up to $E_{\text{F}}$, is almost constant of $\sim1.20$ upon doping.


\bibitem{CaCuO2}
    Electronic structure of an isostructural CaCuO$_{2}$
    is also computed for comparison.
    From the DMFT valence histogram of CaCuO$_{2}$,
    Hund's physics is not noticeable in the cuprate
    while Mott physics is evident therein
    (see Figs.~S9 and S18 in Supplemental Material~\cite{Suppl}).


\bibitem{Medici2011}
    Luca de'Medici, Jernej Mravlje, and Antoine Georges,
    ``Janus-Faced Influence of Hund's Rule Coupling in Strongly Correlated Materials,''
    \href{https://doi.org/10.1103/PhysRevLett.107.256401}
    {Phys. Rev. Lett. {\bf 107}, 256401 (2011)}.


\bibitem{Georges2013}
    Antoine Georges, Luca de' Medici, and Jernej Mravlje,
    ``Strong Correlations from Hund's Coupling,''
    \href{https://doi.org/10.1146/annurev-conmatphys-020911-125045}
    {Annual Review of Condenced Matter Physics {\bf 4}, 137 (2013)}.





\end{thebibliography}

\begin{thebibliography}{99}
\bibitem{Wien2k}
    Peter Blaha, Karlheinz Schwarz, Fabien Tran, Robert Laskowski, Georg K. H. Madsen, and Laurence D. Marks,
    ``WIEN2k: An APW+lo program for calculating the properties of solids,''
    \href{https://doi.org/10.1063/1.5143061}
    {J. Chem. Phys. {\bf 152}, 074101 (2020)}.

\bibitem{Haule2010-prb}
    Kristjan Haule, Chuck-Hou Yee, and Kyoo Kim,
    ``Dynamical mean-field theory within the full-potential methods: Electronic structure of CeIrIn$_{5}$, CeCoIn$_{5}$, and CeRhIn$_{5}$,''
    \href{https://doi.org/10.1103/PhysRevB.81.195107}
    {Phys. Rev. B {\bf 81}, 195107 (2010)}.

\bibitem{Hayward1999-jacs}
    M. A. Hayward, M. A. Green, M. J. Rosseinsky, and J. Sloan,
    ``Sodium Hydride as a Powerful Reducing Agent for Topotactic Oxide Deintercalation: Synthesis and Characterization of the Nickel(I) Oxide LaNiO$_{2}$,''
    \href{https://doi.org/10.1021/ja991573i}
    {J. Am. Chem. Soc. {\bf 121}, 8843 (1999)}.

\bibitem{Werner2006}
    Philipp Werner, Armin Comanac, Luca de' Medici, Matthias Troyer, and Andrew J. Millis,
    ``Continuous-Time Solver for Quantum Impurity Models,''
    \href{https://doi.org/10.1103/PhysRevLett.97.076405}
    {Phys. Rev. Lett. {\bf 97}, 076405 (2006)}.

\bibitem{Haule2007-ctqmc}
    Kristjan Haule,
    ``Quantum Monte Carlo impurity solver for cluster dynamical mean-field theory and electronic structure calculations with adjustable cluster base,''
    \href{https://doi.org/10.1103/PhysRevB.75.155113}
    {Phys. Rev. B {\bf 75}, 155113 (2007)}.

\bibitem{Perdew1996}
    John P. Perdew, Kieron Burke, and Matthias Ernzerhof,
    ``Generalized Gradient Approximation Made Simple,''
    \href{https://doi.org/10.1103/PhysRevLett.77.3865}
    {Phys. Rev. Lett. {\bf 77}, 3865 (1996)}.

\bibitem{Haule2015}
    Kristjan Haule,
    ``Exact Double Counting in Combining the Dynamical Mean Field Theory and the Density Functional Theory,''
    \href{https://doi.org/10.1103/PhysRevLett.115.196403}
    {Phys. Rev. Lett. {\bf 115}, 196403 (2015)}.

\bibitem{Jarrell1996-maxent}
    Mark Jarrell, J. E. Gubernatis,
    ``Bayesian inference and the analytic continuation of imaginary-time quantum Monte Carlo data,''
    \href{https://doi.org/10.1016/0370-1573(95)00074-7}
    {Physics Reports {\bf 269}, 133 (1996)}.



\bibitem{Pickett2020}
    Mi-Young Choi, Kwan-Woo Lee, and Warren E. Pickett,
    ``Role of $4f$ states in infinite-layer NdNiO$_{2}$''
    \href{https://doi.org/10.1103/PhysRevB.101.020503}
    {Phys. Rev. B {\bf 101}, 020503(R) (2020)}.


\bibitem{Qazilbash2009}
    M. M. Qazilbash, J. J. Hamlin, R. E. Baumbach, Lijun Zhang, D. J. Singh, M. B. Maple, and D. N. Basov,
    ``Electronic correlations in the iron pnictides,''
    \href{https://doi.org/10.1038/nphys1343}
    {Nature Physics {\bf 5}, 647 (2009)}.


\bibitem{Deng2014}
    Xiaoyu Deng, Aaron Sternbach, Kristjan Haule, D. N. Basov, and Gabriel Kotliar,
    ``Shining Light on Transition-Metal Oxides: Unveiling the Hidden Fermi Liquid,''
    \href{https://doi.org/10.1103/PhysRevLett.113.246404}
    {Phys. Rev. Lett. {\bf 113}, 246404 (2014)}.


\bibitem{Mravlje2011}
    Jernej Mravlje, Markus Aichhorn, Takashi Miyake, Kristjan Haule, Gabriel Kotliar, and Antoine Georges,
    ``Coherence-Incoherence Crossover and the Mass-Renormalization Puzzles in Sr$_{2}$RuO$_{4}$,''
    \href{https://doi.org/10.1103/PhysRevLett.106.096401}
    {Phys. Rev. Lett. {\bf 106}, 096401 (2011)}.


\bibitem{Mravlje2016}
    Jernej Mravlje and Antoine Georges,
    ``Thermopower and Entropy: Lessons from Sr$_{2}$RuO$_{4}$,''
    \href{https://doi.org/10.1103/PhysRevLett.117.036401}
    {Phys. Rev. Lett. {\bf 117}, 036401 (2016)}.


\bibitem{Haule2009-njp}
    K. Haule and G. Kotliar,
    ``Coherence-incoherence crossover in the normal state of iron oxypnictides and importance of Hund's rule coupling,''
    \href{https://doi.org/10.1088/1367-2630/11/2/025021}
    {New J. Phys. {\bf 11}, 025021 (2009)}.


\bibitem{Hardy2013}
    F. Hardy, A. E. B\"{o}hmer, D. Aoki, P. Burger, T. Wolf, P. Schweiss, R. Heid, P. Adelmann, Y. X. Yao, G. Kotliar, J. Schmalian, and C. Meingast,
    ``Evidence of Strong Correlations and Coherence-Incoherence Crossover in the Iron Pnictide Superconductor KFe$_{2}$As$_{2}$,''
    \href{https://doi.org/10.1103/PhysRevLett.111.027002}
    {Phys. Rev. Lett. {\bf 111}, 027002 (2013)}.


\bibitem{Miao2016}
    H. Miao, Z. P. Yin, S. F. Wu, J. M. Li, J. Ma, B.-Q. Lv, X. P. Wang, T. Qian, P. Richard, L.-Y. Xing, X.-C. Wang, C. Q. Jin, K. Haule, G. Kotliar, and H. Ding,
    ``Orbital-differentiated coherence-incoherence crossover identified by photoemisson spectroscopy in LiFeAs,''
    \href{https://doi.org/10.1103/PhysRevB.94.201109}
    {Phys. Rev. B {\bf 94}, 201109(R) (2016)}.


\bibitem{Liu2015}
    Z. K. Liu, M. Yi, Y. Zhang, J. Hu, R. Yu, J.-X. Zhu, R.-H. He, Y. L. Chen, M. Hashimoto, R. G. Moore, S.-K. Mo, Z. Hussain, Q. Si, Z. Q. Mao, D. H. Lu, and Z.-X. Shen,
    ``Experimental observation of incoherent-coherent crossover and orbital-dependent band renormalization in iron chalcogenide superconductors,''
    \href{https://doi.org/10.1103/PhysRevB.92.235138}
    {Phys. Rev. B {\bf 92}, 235138 (2015)}.



\bibitem{wannier90-1}
    Arash A. Mostofi, Jonathan R. Yates, Young-Su Lee, Ivo Souza, David Vanderbilt, Nicola Marzari,
    ``wannier90: A tool for obtaining maximally-localised Wannier functions,''
    \href{https://doi.org/10.1016/j.cpc.2007.11.016}
    {Computer Physics Communications {\bf 178}, 685 (2008)}.


\bibitem{wannier90-2}
    Giovanni Pizzi {\it et al.},
    ``Wannier90 as a community code: new features and applications,''
    \href{https://doi.org/10.1088/1361-648X/ab51ff}
    {J. Phys.: Condens. Matter {\bf 32}, 165902 (2020)}.


\bibitem{Wan2020-arXiv}
    Xiangang Wan, Vsevolod Ivanov, Giacomo Resta, Ivan Leonov, Sergey Y. Savrasov,
    ``Calculated Exchange Interactions and Competing $S$ = 0 and $S$ = 1 States in Doped NdNiO$_{2}$''
    \href{https://arxiv.org/abs/2008.07465}
    {arXiv:2008.07465 (2020)}.


\bibitem{Sakakibara20-prl-suppl}
    Hirofumi Sakakibara, Hidetomo Usui, Katsuhiro Suzuki, Takao Kotani, Hideo Aoki, and Kazuhiko Kuroki,
    ``Model Construction and a Possibility of Cupratelike Pairing in a New $d^9$ Nickelate Superconductor (Nd,Sr)NiO$_{2}$''
    \href{https://doi.org/10.1103/PhysRevLett.125.077003}
    {Phys. Rev. Lett. {\bf 125}, 077003 (2020)}.


\bibitem{Nomura2019-prb-suppl}
    Yusuke Nomura, Motoaki Hirayama, Terumasa Tadano, Yoshihide Yoshimoto, Kazuma Nakamura, and Ryotaro Arita,
    ``Formation of a two-dimensional single-component correlated electron system and band engineering in the nickelate superconductor NdNiO$_{2}$'',
    \href{https://doi.org/10.1103/PhysRevB.100.205138}
    {Phys. Rev. B {\bf 100}, 205138 (2019)}.


\bibitem{Lee2020-nm-suppl}
    M. Hepting, D. Li, C. J. Jia, H. Lu, E. Paris, Y. Tseng, X. Feng, M. Osada, E. Been, Y. Hikita, Y.-D. Chuang, Z. Hussain, K. J. Zhou, A. Nag, M. Garcia-Fernandez, M. Rossi, H. Y. Huang, D. J. Huang, Z. X. Shen, T. Schmitt, H. Y. Hwang, B. Moritz, J. Zaanen, T. P. Devereaux, and W. S. Lee,
    ``Electronic structure of the parent compound of superconducting infinite-layer nickelates,''
    \href{https://doi.org/10.1038/s41563-019-0585-z}
    {Nature Materials {\bf 19}, 381 (2020)}.


\bibitem{Adhikary2020-prb-suppl}
    Priyo Adhikary, Subhadeep Bandyopadhyay, Tanmoy Das, Indra Dasgupta, and Tanusri Saha-Dasgupta,
    ``Orbital-selective superconductivity in a two-band model of infinite-layer nickelates'',
    \href{https://doi.org/10.1103/PhysRevB.102.100501}
    {Phys. Rev. B {\bf 102}, 100501(R) (2020)}.


\bibitem{Hwang2019}
    Danfeng Li, Kyuho Lee, Bai Yang Wang, Motoki Osada, Samuel Crossley, Hye Ryoung Lee, Yi Cui, Yasuyuki Hikita, and Harold Y. Hwang,
    ``Superconductivity in an infinite-layer nickelate,''
    \href{https://doi.org/10.1038/s41586-019-1496-5}
    {Nature {\bf 572}, 624 (2019)}.

\bibitem{Zeng2020-arXiv}
    Shengwei Zeng, Chi Sin Tang, Xinmao Yin, Cgangjian Li, Zhen Huang, Junxiong Hu, Wei Liu, Ganesh Ji Omar, Hariom Jani, Zhi Shiuh Lim, Kun Han, Dongyang Wan, Ping Yang, Andrew T. S. Wee, Ariando Ariando,
    ``Phase diagram and superconducting dome of infinte-layer Nd$_{1-x}$Sr$_{x}$NiO$_{2}$ thin films,''
    \href{https://doi.org/10.1103/PhysRevLett.125.147003}
    {Phys. Rev. Lett. {\bf 125}, 147003 (2020)}.

\bibitem{Li2020-arXiv}
    Danfeng Li, Bai Yang Wang, Kyuho Lee, Shannon P. Harvey, Motoki Osada, Berit H. Goodge, Lena F. Kourkoutis, Harold Y. Hwang,
    ``Superconducting Dome in Nd$_{1-x}$Sr$_{x}$NiO$_{2}$ Infinite Layer Films,''
    \href{https://doi.org/10.1103/PhysRevLett.125.027001}
    {Phys. Rev. Lett. {\bf 125}, 027001 (2020)}.

\end{thebibliography}
\end{document}